\documentclass[aps,prd,reprint,superscriptaddress,nofootinbib]{revtex4-2}

\usepackage{comment}
\usepackage[utf8]{inputenc} 
\usepackage{graphicx,overpic,mathtools}
\usepackage{amsthm,amsmath,amssymb,hyperref,mathrsfs}
\usepackage{braket,bm,bbm,setspace}
\usepackage[normalem]{ulem} 
\usepackage{physics}
\usepackage[makeroom]{cancel}
\hypersetup{
    colorlinks=false,
    pdfborder={0 0 0},
}
\usepackage[usenames,dvipsnames]{xcolor}

\newcommand{\bfk}{\textbf{k}}

\begin{document}

\title{Generalized quantum Vlasov equation for particle creation and unitary dynamics}

\author{\'Alvaro \'Alvarez-Dom\'inguez}
\email{alvalv04@ucm.es}
\affiliation{Departamento de F\'{\i}sica Te\'orica and IPARCOS, Universidad Complutense de Madrid, 28040 Madrid, Spain}  

\author{Luis J. Garay} 
\email{luisj.garay@ucm.es}\affiliation{Departamento de F\'{\i}sica Te\'orica and IPARCOS, Universidad Complutense de Madrid, 28040 Madrid, Spain}  
\affiliation{Instituto de Estructura de la Materia (IEM-CSIC), Serrano 121, 28006 Madrid, Spain}

\author{Mercedes Mart\'in-Benito}
\email{m.martin.benito@ucm.es}
\affiliation{Departamento de F\'{\i}sica Te\'orica and IPARCOS, Universidad Complutense de Madrid, 28040 Madrid, Spain}    

\begin{abstract}
The loss of time-translational invariance caused by a time-dependent external agent leads to particle creation effects in quantum field theory. This phenomenon results in ambiguities when selecting the quantum vacuum of the canonical quantization. In this work we analyze how the time evolution of the number of created particles depends on these ambiguities when external agents are spatially homogeneous. In particular, we generalize the standard quantum Vlasov equation in order to accommodate in its formulation the possibility of having different choices of vacuum, including adiabatic vacua. This study leads us to propose a new physical criterion stronger than the unitary implementation of the dynamics in order to reduce the quantization ambiguities.
\end{abstract}

\maketitle

\section{Introduction}
\label{sec_introduction}

In quantum field theory, particle creation effects exist in many different settings including the Schwinger effect \cite{Sauter1931742,Schwinger1951} in quantum electrodynamics, the Hawking effect \cite{BlackHexplosons} in black holes, and particle creation effects in cosmology \cite{PhysRev.183.1057} such as the Gibbons-Hawking effect~\cite{ PhysRevD.15.2738}. In order to study these phenomena, it is common to consider the dynamics of canonically quantized matter fields coupled to time-dependent external electromagnetic and/or gravitational fields. In such analyses, one usually neglects backreaction, thus assuming a mean-field approximation for the test matter fields, which propagate on a classical background. We will do so in this work. In particular, we will focus our study on the time dependence of the number of created particles $N(t)$ generated in this kind of process.

Particle creation is rooted in the breaking of symmetries caused by external agents. Free fields in flat spacetime possess Poincaré symmetry. When imposing the invariance of the canonical quantum theory under this group of symmetry, there is only one possible basis of modes in which solutions to the equations of motion can be expanded: plane waves. This basis determines unique sets of annihilation and creation operators, which in turn define the Fock vacuum of the quantum theory: the so-called Minkowski vacuum. However, for instance, when a time-dependent external field is coupled to matter fields, the classical Hamiltonian is no longer invariant under time translations and there is freedom in the choice of the annihilation and creation operators (and thus,  the vacuum) of the corresponding Fock quantization, which  is therefore not unique. Depending on the choice of vacuum at each time, its evolution might produce particle-antiparticle pairs, determining the time dependence of $N(t)$.   Due to these ambiguities, the physical interpretation of $N(t)$ and other physical observables such as the energy density is an ongoing discussion~\cite{superadiabatic1,superadiabatic2,IldertonZerothOrder,Yamada_2021}.

There can be found different choices of vacuum in the literature. In general, its selection depends on the particular system under study and on the properties that we want to impose to its quantum theory. One of the most common options is to choose adiabatic vacua, introduced by Parker in \cite{PhysRev.183.1057} and later formalized by Lüders and Roberts in \cite{LudersRoberts}. They are intensively used not only in cosmology but also in the context of the Schwinger effect~\cite{Kluger_1998,MottolaDeSitter99}. These adiabatic modes are understood to be the most natural and simple extensions of Minkowski plane waves when the external agent slowly varies throughout time. However, many other options motivated by different criteria are also legitimate \emph{a priori}. Some examples so far considered are: modes diagonalizing the Hamiltonian in cosmology \cite{fahn2019dynamical,HamiltonianDiagonalizationGuillermo,Cortez:2020rla}, modes minimizing oscillations in the primordial power spectrum in cosmology \cite{PrimordialPowerJavi,nonoscillations},  adiabatic  modes minimizing oscillations in the number of created particles by the Schwinger effect and   by cosmological settings  \cite{superadiabatic1,superadiabatic2,Yamada_2021}, etc. In this work, we will be interested in generalizing standard expressions found in the literature for the evolution of the particle number $N(t)$ by considering arbitrary selections of modes for the quantization. Bogoliubov transformations of the canonical quantization approach will allow us to address this question.

In the study of classical non-equilibrium physical systems, kinetic theory  has been a very successful tool \cite{liboff}. In particular, when describing a system composed by identical particles, the starting point in this theory is the Liouville equation for the joint probability distribution of the entire system. If we assume that particles are weakly correlated, we can deduce a closed equation of motion for the probability distribution of each individual particle: the so-called classical Vlasov equation. This equation does not consider collisions between particles. This can be accomplished with a more general but complicated approximation: the Boltzmann kinetic equation. A generalization to quantum field theory of the classical Vlasov equation should contemplate particle creation. This is done in the context of the quantum kinetic approach. The widely accepted proposal, based on incorporating a particle creation term, is the so-called quantum Vlasov equation (QVE), which is an integro-differential equation for $N(t)$. In the context of the Schwinger effect, this equation was first presented in \cite{Kluger_1998} for scalar charged fields under a spatially homogeneous and time-dependent external electric field. Later, its extension to fermionic quantum fields was proposed in \cite{1998}. This equation and its formalism has been used in a wide range of frameworks, including continuum strong QCD \cite{RobertsSchmidt}, electron-positron pair creation in QED (from nuclei phenomena to black hole physics)~\cite{Ruffini}, laser technology~\cite{Dunne,DumluDunne,HebenstreitShortLaserPulses}, or in cosmology considering a de Sitter spacetime \cite{MottolaDeSitter99,MottolaDeSitter}. Most of the literature use this QVE for the particular choice of vacuum defined by zeroth-order adiabatic modes. One of the main aims of this work is to generalize this equation to arbitrary vacua, getting in this way a ``generalized QVE''. Later we will restrict this generalized QVE to higher order adiabatic vacua \cite{MottolaDeSitter99}. 

In order to reduce the ambiguities in the quantization of classical theories with no time-translational invariance, a criterion has been proposed in various settings, from homogeneous cosmologies \cite{CORTEZ201536,Cortez:2019orm,Cortez:2020rla,universe7080299} to the Schwinger effect \cite{Garay2020,AlvarezFermions}: the unitary implementation in the quantum theory of matter fields dynamics. Weaker conditions are also found in the literature, imposing that only the in and out states (at asymptotic past and future times) are related by a unitary $S$-matrix \cite{Gavrilov:1996pz,WALD1979490}. The motivation for imposing the former stronger requirement, at all intermediate times, is two-fold. First, we ensure that quantum theories at all times are physically equivalent, in the sense that they provide the same probability amplitudes. Moreover, in those references it was proved that in a wide range of settings this requirement reduces the ambiguities in the quantization to a unique family of unitarily equivalent quantizations. Second, it ensures that the total number of created particles is well defined (i.e., finite) at every finite time. 

The unique family of vacua associated with the quantizations that unitarily implement the dynamics is precisely the family to which we will restrict our previously found generalized QVE.  We will see that there is an interesting connection between the usual QVE and its generalization to modes unitarily implementing the dynamics: under certain conditions, the former is precisely the leading order of the latter in the ultraviolet regime.  This will allow us to propose a more strict criterion for reducing the ambiguity in the quantization based on the ultraviolet behavior of the generalized QVE. 

For definiteness, we will consider a charged scalar field in the presence of a spatially homogeneous but time-dependent electric field, system studied in \cite{Garay2020}, although extensions to other homogeneous systems follow straightforwardly.

The structure of the paper is as follows.
In section \ref{sec_schwingercurved} we specify the pair creation effects to which the generalizations of the expressions found in the literature for $N(t)$ might be extended. In section \ref{sec_canonical} we present the key ideas of the canonical quantization approach, parametrizing the ambiguities and deducing a general expression for $N(t)$. These results will be used in section \ref{sec_GQVE}, where we will obtain the generalization to arbitrary quantizations of the quantum Vlasov equation. In section \ref{sec_unitary}, we will analyze the unitary dynamics criterion in the scalar Schwinger effect, particularizing our generalized QVE to modes satisfying this requirement.  We will also propose an additional criterion for reducing the quantization ambiguities. Finally, section \ref{sec_conclusions} is devoted to summarizing and discussing the essential results of this work.

\section{From the Schwinger effect to curved spacetimes} 
\label{sec_schwingercurved}

In order to study the Schwinger effect, let us consider a scalar field $\phi(t,\textbf{x})$ of mass $m$ and charge $q$ propagating in Minkowski spacetime under an external time-dependent electric field with four-vector potential $A_{\mu}$. It satisfies the Klein-Gordon equation of motion
\begin{equation} \label{eq_KG}
    \left[ (\partial_{\mu}+iqA_{\mu})(\partial^{\mu}+iqA^{\mu})+m^2 \right]\phi(t,\textbf{x})=0.
\end{equation}
In this work we will also assume that the electric field is homogeneous, although not necessarily isotropic. We will use the temporal gauge, i.e., $A_{\mu}(t)=(0,\textbf{A}(t))$. Therefore, after Fourier transforming \eqref{eq_KG} the time-dependent $\bfk$-modes
\begin{equation} \label{eq_Fourier}
    \phi_{\bfk}(t)=\int \frac{d^3\textbf{x}}{(2\pi)^{3/2}} \ e^{-i\bfk\cdot\textbf{x}}\phi(t,\textbf{x})
\end{equation}
satisfy decoupled harmonic oscillator equations
\begin{equation} \label{eq_harmonic} 
    \ddot{\phi}_{\bfk}(t)+\omega_{\bfk}(t)^2\phi_{\bfk}(t)=0,
\end{equation}
with time-dependent frequencies
\begin{equation} \label{eq_omegaSchwinger}
    \omega_{\bfk}(t)=\sqrt{[\bfk+q\textbf{A}(t)]^2+m^2}.
\end{equation}
Note that complex scalar modes $\phi_{\bfk}(t)$ can be split into their real and imaginary parts, both satisfying harmonic oscillator equations \eqref{eq_harmonic}. Thus, from now on we will consider without loss of generality $\phi_{\bfk}(t)$ as real variables.

Observe that the dependence on the electric field only appears in $\omega_{\bfk}(t)$. Due to the fact that we are treating it as an external agent, frequencies $\omega_{\bfk}(t)$ are fixed and not affected by the dynamics of the modes $\phi_{\bfk}(t)$. In other words, we neglect backreaction effects, only dealing with \eqref{eq_harmonic} and forgetting about the equation of motion of the external field.

Although our working example will be the scalar Schwinger effect, our approach can be easily extended to many other systems. Indeed, for most parts of this work we are not going to use the explicit expression of the Schwinger frequency \eqref{eq_omegaSchwinger}, except for section~\ref{sec_unitary}, where we will use a system-dependent reasoning. Thus, the key requirement that a theory has to verify so that our formalism is applicable is that it can be characterized by a collection of real degrees of freedom satisfying decoupled harmonic oscillator equations with time-dependent frequencies. 

For instance, we can consider systems that have degrees of freedom $\psi_i(t)$ verifying the equation of motion of a damped oscillator
\begin{equation} \label{eq_eom2}
    \ddot{\psi}_i(t)+2\gamma_i(t)\dot{\psi}_i(t)+\Omega_i(t)^2\psi_i(t)=0,
\end{equation}
since there always exists a canonical transformation $\psi_i(t)=e^{-\int^t dt' \gamma_i(t')}\phi_i(t)$ which removes the first order term \cite{doi:10.1063/1.527707}, transforming the equation of motion \eqref{eq_eom2} for $\psi_i(t)$ into a harmonic oscillator equation for $\phi_i(t)$ with time-dependent frequency
\begin{equation}
    \omega_i(t)=\sqrt{\Omega_i(t)^2- \dot{\gamma}_i(t)-\gamma_i(t)^2}.
\end{equation} 

A second example of a system characterized by equations of the type \eqref{eq_harmonic} is a fermionic field coupled to a homogeneous time-dependent electric field. In this case, the Fourier transform of the Dirac equation yields fermionic modes formed by four real variables satisfying~\eqref{eq_harmonic} with time-dependent frequencies (see e.g. \cite{1998,AlvarezFermions})
\begin{equation}
    \omega_{\bfk}^{(\pm)}(t)=\sqrt{[\bfk+q\textbf{A}(t)]^2+m^2\pm iq|\dot{\textbf{A}}(t)|}.
\end{equation}
However, these variables are not completely decoupled and minor manipulations inspired in references \cite{1998,AlvarezFermions} should be applied to the procedure followed here in order to extend these results to the fermionic case.

Other significant examples are found in cosmological settings. Let $\varphi$ be a real scalar field with mass $m$ in a Friedmann-Lemaître-Robertson-Walker (FLRW) spacetime, defined by the well-known metric 
\begin{equation}
    ds^2=a(t)^2\left(-dt^2+h_{ij}dx^{i}dx^{j}\right).
\end{equation}
Here $a$ is the scale factor, $t$ is the conformal time, and $h_{ij}$ is the time-independent three-dimensional metric on a spatial hypersurface~$\Sigma$. It can be easily seen \cite{Cortez:2019orm} that the redefined scalar field $\phi=a\varphi$ satisfies
\begin{equation} \label{eq_cosmologyKG}
    \ddot\phi-\Delta\phi+m(t)^2\phi=0, \quad m(t)=\sqrt{m^2a(t)^2-\ddot a(t)/a(t)},
\end{equation}
where   $\Delta$ is the Laplace-Beltrami operator on the spatial hypersurface~$\Sigma$. While in the Schwinger effect the agent generating particle production is the external electric field, now in FLRW spacetimes the particle production is due to the evolution of the Universe, characterized by $a(t)$. An alternative interpretation is that the field $\phi$ is a free field propagating in the static spacetime $ds^2=-dt^2+h_{ij}dx^{i}dx^{j}$, but with a time-dependent mass $m(t)$. In order to obtain a harmonic oscillator equation of the type \eqref{eq_harmonic} for certain modes, different orthonormal bases for the expansions of the solutions to \eqref{eq_cosmologyKG} can be chosen depending on the particular system. For example, if $\Sigma$ is a three-sphere in a closed FLRW spacetime, an expansion in terms of hyperspherical harmonics of order $n$ leads to decoupled modes satisfying harmonic oscillator equations with time-dependent frequencies \cite{Cortez:2019orm}
\begin{equation} \label{eq_omegan}
    \omega_n(t)=\sqrt{n(n+2)+m(t)^2}.
\end{equation} 

In all these systems the external agent (either the electric or the gravitational field) is assumed to be spatially homogeneous. Thanks to this symmetry, it is possible to find modes of the scalar matter field verifying decoupled harmonic oscillator equations with time-dependent frequencies. However, this is not the case when dealing with spatial inhomogeneities. In that case the mode decomposition would lead to a tower of coupled equations of motion for the infinite modes of the field and we would need other techniques. For example, in the case of the inhomogeneous Schwinger effect, one could consider the Wigner approach \cite{Hebenstreit1,Hebenstreit2,Hebenstreit3,Schwingerwigner,WignerFonarev}. We leave those more complicated systems for future work.

\section{Canonical quantization approach} \label{sec_canonical}

In this section we will quantize the classical systems described in the previous section following the canonical quantization approach. We present here the essential aspects in order to understand our work. For deeper analyses covering a wide range of systems of the type described in section \ref{sec_schwingercurved}, see e.g. \cite{wald1994quantum,Garay2020,AlvarezFermions,Cortez:2019orm, Cortez:2020rla}.

\subsection{Ambiguities in the canonical quantization}

Given a particular complex solution $z_{\bfk}(t)$ of the harmonic oscillator equation with time-dependent frequency \eqref{eq_harmonic}, 
there exists a unique complex coefficient $a_{\bfk}$ such that 
any other real solution $\phi_{\bfk}(t)$ and its canonically conjugate momentum $\pi_{\bfk}(t)=\dot{\phi}_{\bfk}(t)$ can be uniquely written as
\begin{align} 
    \mqty(\phi_{\bfk}(t)\\\pi_{\bfk}(t))&=G_{(z_{\bfk},\dot z_{\bfk})}(t)\mqty(a_{\bfk}\\a^*_\bfk), 
    \notag\\
     G_{(z_{\bfk},\dot z_{\bfk})}(t)&=
    \mqty(z_{\bfk}(t)&z_{\bfk}^*(t)\\\dot z_{\bfk}(t)& \dot z_{\bfk}^*(t)).
\label{eq_matrixz}\end{align}
The coefficient $a_{\bfk}$ and its complex conjugate $a_{\bfk}^*$  of this linear combination  are called annihilation and creation variables, respectively.

The above decomposition  depends on the choice of  the solution  $z_\bfk(t)$. More generally,  we have the possibility of expressing the solution $\phi_{\bfk}(t)$ and its momentum $\pi_{\bfk}(t)$ in terms of complex functions $\zeta_{\bfk}(t)$ and $\rho_{\bfk}(t)$, respectively:\footnote{In the literature about the canonical study of the quantum unitary implementation of the dynamics (e.g., see references \cite{Garay2020,AlvarezFermions,Cortez:2019orm}) it is usual to denote these time-dependent functions as $\zeta_{\bfk}=ig_{\bfk}^*$ and $\rho_{\bfk}=-if_{\bfk}^*$.}
\begin{align} 
    \mqty(\phi_{\bfk}(t)\\\pi_{\bfk}(t))&=G_{(\zeta_{\bfk},\rho_{\bfk})}(t)\mqty(a_{\bfk}(t)\\a^*_\bfk(t)), \notag\\ G_{(\zeta_{\bfk},\rho_{\bfk})}(t)&=
    \mqty(\zeta_{\bfk}(t)&\zeta_{\bfk}^*(t)\\\rho_{\bfk}(t)& \rho_{\bfk}^*(t)).
\label{eq_matrix}\end{align}
Let us remark that $\zeta_{\bfk}(t)$ is not necessarily a solution to the harmonic oscillator equation~\eqref{eq_harmonic}. Only if this is the case, the annihilation and creation variables $a_{\bfk}(t)$ and $a^*_{\bfk}(t)$ are time-independent. Otherwise, these variables have to carry the appropriate time dependence compensating for that of $\zeta_{\bfk}(t)$ and $\rho_{\bfk}(t)$, so that the combination \eqref{eq_matrix} leads to a solution $\phi_{\bfk}(t)$ of~\eqref{eq_harmonic}. 

Note that equation \eqref{eq_matrix} reduces to \eqref{eq_matrixz}
if we choose $\zeta_{\bfk}(t)=z_{\bfk}(t)$, which then implies $\rho_{\bfk}(t)=\dot{z}_{\bfk}(t)$ (and $a_{\bfk}(t)=a_{\bfk})$. In the general case, i.e., when $\zeta_{\bfk}(t)$ is not a solution to~\eqref{eq_harmonic}, $\zeta_{\bfk}(t)$ and $\rho_{\bfk}(t)$ are not completely independent. Indeed, the pair of canonical modes and the annihilation and creation variables have to verify the Poisson bracket relations
\begin{align}
    \{\phi_{\bfk}(t),\pi_{\bfk'}(t)\}&=\delta(\bfk-\bfk'), \notag\\ \{a_{\bfk}(t),a_{\bfk'}^*(t)\}&=-i\delta(\bfk-\bfk'),
\end{align}
where $\delta$ denotes the Dirac delta distribution. They impose the normalization conditions
\begin{equation} \label{eq_relzetarho}
    \zeta_{\bfk}(t)\rho_{\bfk}^*(t)-\zeta_{\bfk}^*(t)\rho_{\bfk}(t)=i.
\end{equation}
It can be easily verified that this requirement ensures that the expression for $\pi_{\bfk}(t)$ given 
 in the second row of \eqref{eq_matrix} is equivalent to the time derivative of $\phi_{\bfk}(t)$ in the first row.

In the canonical quantization approach we promote the annihilation and creation variables to annihilation and creation operators acting on the corresponding Fock space. Then, in view of \eqref{eq_matrix}, one classical theory can have infinitely many associated quantum theories. Indeed,   we have the ambiguity in the particular choice of functions  $(\zeta_{\bfk}(t),\rho_{\bfk}(t))$,  which have to verify the relation \eqref{eq_relzetarho}. This selection determines a one-parameter family of quantizations, one for each value of the time variable~$t$: the corresponding quantum operators $(\hat{a}_{\bfk}(t),\hat{a}_{\bfk}(t)^{\dagger})$ determine the associated Fock vacuum state $|0\rangle_t$ as the state annihilated by $\hat{a}_{\bfk}(t)$ for all $\bfk$. In other words, and connecting with analog discussions in the literature of unitary implementation of the quantum field dynamics (see e.g. \cite{CORTEZ201536}), we have ambiguity in the choice of canonical variables to be quantized  and in the choice of complex structure to carry out the quantization, both encoded   in  the functions $(\zeta_{\bfk}(t),\rho_{\bfk}(t))$.

One criterion to reduce these ambiguities is to unitarily implement the symmetries of the classical system in the quantum theory, which reduces the possible selections of functions $(\zeta_{\bfk}(t),\rho_{\bfk}(t))$. In fact, in Minkowski spacetime when no external field is present, Poincaré symmetry fixes completely this choice: this maximal symmetry fixes $\zeta_{\bfk}(t)$ to be the plane wave of frequency $\omega_{\bfk}$. For systems which only differ slightly from flat spacetime, one can expect that this construction can be extended. This is the case when $\omega_{\bfk}(t)$ varies slowly throughout time, recovering the Minkowski case in the limit of constant frequency. However, in our work we will go beyond this particular case.  

Let us note that in our system, requiring that our quantization unitarily implements the classical symmetries implies invariance of the vacuum under spatial translations. As a consequence, other expansions of the canonical pair $(\phi(t,\textbf{x}),\pi(t,\textbf{x}))$ mixing Fourier modes are not allowed, and all the ambiguity that we are considering is the one encoded in the choice of $ (\zeta_{\bfk}(t),\rho_{\bfk}(t))$.

\subsection{Parametrization of the ambiguities} \label{sec_parametrization}

For later convenience, we parametrize the freedom in the choice of $\zeta_{\bfk}(t)$ in terms of two arbitrary real functions $W_{\bfk}(t) > 0$ and $\varphi_{\bfk}(t)$ related to its modulus and its phase, respectively, in the following way:
\begin{equation} \label{eq_zetaparam}
    \zeta_{\bfk}(t)=\frac{1}{\sqrt{2W_{\bfk}(t)}}e^{-i\varphi_{\bfk}(t)}.
\end{equation}
In addition, it is easy to verify that the normalization condition \eqref{eq_relzetarho} reduces the ambiguity in the choice of the complex function $\rho_{\bfk}(t)$ to just one real function $Y_{\bfk}(t)$ such that
\begin{equation} \label{eq_rhoparam}
    \rho_{\bfk}(t)=-\sqrt{\frac{W_{\bfk}(t)}{2}}[i+Y_{\bfk}(t)]e^{-i\varphi_{\bfk}(t)}.
\end{equation}

There are occasions in which certain families of functions $(\zeta_{\bfk}(t),\rho_{\bfk}(t))$ stand out. In the particular case in which we demand $\zeta_{\bfk}(t)$ to be a solution to the harmonic oscillator equation with time-dependent frequency \eqref{eq_harmonic}, then not only the normalization condition~\eqref{eq_relzetarho} has to be verified but also $\rho_{\bfk}(t)=\dot{\zeta}_{\bfk}(t)$. This fixes $\dot{\varphi}_{\bfk}(t)$ and $Y_{\bfk}(t)$ as functions of $W_{\bfk}(t)$ according to
\begin{align} 
\label{eq_eqW}
    W_{\bfk}(t)^2&=\omega_{\bfk}(t)^2-\frac{1}{2}\left[ \frac{\ddot{W}_{\bfk}(t)}{W_{\bfk}(t)}-\frac{3}{2}\frac{\dot{W}_{\bfk}(t)^2}{W_{\bfk}(t)^2} \right], 
\\ \label{eq_phigammasol}
    \dot{\varphi}_{\bfk}(t)&=W_{\bfk}(t), \qquad Y_{\bfk}(t)=\frac{\dot{W}_{\bfk}(t)}{2W_{\bfk}(t)^2}.
\end{align}
Thus, the freedom in the choice of the pair $(\zeta_{\bfk}(t),\rho_{\bfk}(t))$ when we impose that $\zeta_{\bfk}(t)$ is a particular normalized solution to \eqref{eq_harmonic} is encoded in the initial conditions $W_{\bfk}(t_0)$, $\dot{W}_{\bfk}(t_0)$, and $\varphi_{\bfk}(t_0)$ at some initial time~$t_0$.

Another possibility is to require that $\zeta_{\bfk}(t)$ is an approximate solution to the equation of motion. In this case, equations \eqref{eq_eqW} and \eqref{eq_phigammasol} must hold approximately. For instance, when the time-dependent frequency $\omega_{\bfk}(t)$ is slowly-varying, the most common selection in the literature is the adiabatic approximation  \cite{birrell_davies_1982}. It is recursively defined from the zeroth-order approximation \begin{equation} \label{eq_adiabatic0}
    W_{\bfk}^{(0)}(t)=\omega_{\bfk}(t), \qquad \varphi^{(0)}_{\bfk}(t)=\int_{t_0}^t dt' \ \omega_{\bfk}(t').
\end{equation}
For $Y^{(0)}_{\bfk}(t)$ there are two choices with different adiabatic order (i.e., number of time derivatives of $\omega_{\bfk}(t)$). The first possibility $Y^{(0)}_\bfk (t)=\dot\omega_\bfk(t)/[2\omega_\bfk(t)^2]$ is common in the references about quantum field theory in curved spacetime (see, e.g., \cite{birrell_davies_1982}). This approximates exact modes (called zeroth-order adiabatic modes) and their derivatives up to second-adiabatic order. The other choice $\breve Y^{(0)}_\bfk(t)=0$ (we have added a $\,\breve{}\,$ to differentiate it from the previous option) is common in the QVE literature (see, e.g., \cite{Kluger_1998,1998,Fedotov_2011}). This approximates exact adiabatic modes only up to first-adiabatic order. We will emphasize the consequences of these two different selections later in the text. The $n$th-adiabatic approximation can be obtained in the standard way \cite{birrell_davies_1982} recursively introducing the previous order in \eqref{eq_eqW}.
The corresponding exact mode $z_{\bfk}^{(n)}(t)$, determined by fixing the initial data according to $z_{\bfk}^{(n)}(t_0)=\zeta^{(n)}_{\bfk}(t_0)$ and $\dot{z}_{\bfk}^{(n)}(t_0)=\rho^{(n)}_{\bfk}(t_0)$, is usually called the $n$th-order adiabatic mode.\footnote{In the context of cosmology it is also usual to define the zeroth-order adiabatic approximation via $W^{(0)}_{\bfk}(t)=k$ and $Y^{(0)}_{\bfk}(t)=0$ \cite{parker_toms_2009}. With that convention the $(n+2)$th-order adiabatic mode is our $n$th-order adiabatic mode.}

On the other hand, remember that in general  we do not require $(\zeta_{\bfk}(t),\rho_{\bfk}(t))$ to be solutions to the equation of motion, not even approximately. We find in the literature other selections, including functions diagonalizing the Hamiltonian for large wave numbers \cite{HamiltonianDiagonalizationGuillermo,Cortez:2020rla}, and others which minimize oscillations of the number of created particles throughout time or of the primordial power spectrum \cite{superadiabatic1,superadiabatic2,PrimordialPowerJavi,nonoscillations}. Moreover, recently the so-called exact WKB analysis has been used, which consists of a Borel resummation of the ordinary WKB approximations, to study the Schwinger effect \cite{exactWKB}. Our analysis will be general, without assuming specific selections of these functions. In section~\ref{sec_unitary} we will restrict the study to the family of Fock quantizations with unitary dynamics, as we consider that property as essential.

In summary, a particular family of  canonical quantum theories  (one for each time $t$) is unequivocally selected by choosing $(\zeta_{\bfk}(t),\rho_{\bfk}(t))$ for each $\bfk$. The preservation of the Poisson algebra of the canonical fields and the creation and annihilation variables at each time restricts in a precise way the choice of $\rho_{\bfk}(t)$ through the normalization condition~\eqref{eq_relzetarho}. We have a complete freedom of two real time-dependent functions ($W_{\bfk}(t)$, $\varphi_{\bfk}(t)$) to determine $\zeta_{\bfk}(t)$ and only one additional real function $Y_{\bfk}(t)$ to characterize $\rho_{\bfk}(t)$. As we are going to see, the number of created particles throughout time will strongly depend on the choice of both $W_{\bfk}(t)$ and $Y_{\bfk}(t)$.

\subsection{Number of created particles} \label{sec_N(t)}

We are interested in computing the number of particles in the vacua $|0\rangle_t$ with respect to the vacuum of another quantum theory that we will take as reference. Furthermore, we will see the variation in $t$ in the functions $(\zeta_{\bfk}(t),\rho_{\bfk}(t))$ as providing time evolution for the quantization, so that particles are created or destroyed as time evolves.  For this comparison, first we have to choose such reference vacuum. With that aim we fix a complex basis $(z_{\bfk}(t), z^*_{\bfk}(t))$ for the space of  solutions of the harmonic oscillator equation \eqref{eq_harmonic}, and determine the associated creation and annihilation time-independent variables $({a}_{\bfk}, {a}^*_{\bfk})$. Then, the reference vacuum, that will be denoted by~$\ket{0}$, will be the state annihilated by all the operators~$\hat{a}_{\bfk}$.

The different sets of annihilation and creation variables $({a}_{\bfk}, {a}_{\bfk}^*)$ and $({a}_{\bfk}(t), {a}_{\bfk}^*(t))$, associated with $(z_{\bfk}(t), z^*_{\bfk}(t))$ and $(\zeta_{\bfk}(t),\rho_{\bfk}(t))$, respectively, are related by a canonical transformation $\mathcal{B}(t)$ called a Bogoliubov transformation. Since the modes $\phi_{\bfk}(t)$ satisfy decoupled harmonic oscillator equations \eqref{eq_harmonic} for different wave vectors $\bfk$, $\mathcal{B}(t)$ does not mix them. Its $\bfk$-component $\mathcal{B}_{\bfk}(t)$ can be written as
\begin{equation} \label{eq_bogoliubov}
    \mqty(a_{\bfk}(t)\\a_{\bfk}^*(t))=\mathcal{B}_{\bfk}(t)\mqty(a_{\bfk}\\a_{\bfk}^*), \qquad \mathcal{B}_{\bfk}(t)=\mqty(\alpha_{\bfk}(t)&\beta_{\bfk}(t)\\\beta_{\bfk}^*(t)&\alpha_{\bfk}^*(t)).
\end{equation}
The preservation of the Poisson algebra of the creation and annihilation variables relates the Bogoliubov coefficients for all $t$ according to
\begin{equation} \label{eq_relationbog}
    |\alpha_{\bfk}(t)|^2-|\beta_{\bfk}(t)|^2=1.
\end{equation}
This Bogoliubov transformation \eqref{eq_bogoliubov} enables us to better understand the physical consequences of having an ambiguity in the selection of annihilation and creation variables. As long as these $\beta$-coefficients do not vanish, the associated quantum theories will have different notions of particles and antiparticles. 
In this way, the number of particles for each wave vector $\bfk$ in the quantum theory defined by the set $(\hat{a}_{\bfk}(t), \hat{a}_{\bfk}^*(t))$ measured with respect to the reference vacuum $|0\rangle$ is given by
\begin{equation} \label{eq_Nbeta}
    N_{\bfk}(t)=\langle 0|\hat{a}_{\bfk}(t)^{\dagger}\hat{a}_{\bfk}(t)|0\rangle=|\beta_{\bfk}(t)|^2.
\end{equation}
The last equality is obtained by substituting the expression of $\hat{a}_{\bfk}(t)$ in terms of $\hat{a}_{\bfk}$ using~\eqref{eq_bogoliubov}. We see that this number of created particles strongly depends both on the reference vacuum and on the particular functions $(\zeta_{\bfk}(t),\rho_{\bfk}(t))$ chosen, and this will be made explicit in the following.

In order to write an expression for the Bogoliubov coefficients we use the classical equivalence between $\phi_{\bfk}(t)$ and $\pi_{\bfk}(t)$ written in terms of an exact solution $z_{\bfk}(t)$ (equation \eqref{eq_matrixz}) and in terms of $\zeta_{\bfk}(t)$ and $\rho_{\bfk}(t)$ (equation \eqref{eq_matrix}). Using also the normalization condition~\eqref{eq_relzetarho} we finally deduce that
\begin{align}
    \mqty(\alpha_{\bfk}(t)\\\beta_{\bfk}^*(t))&= G_{(\zeta_\bfk,\rho_\bfk)}^{-1}(t) 
    \mqty(z_{\bfk}(t)\\ \dot z_{\bfk}(t)),
    \notag\\
    G_{(\zeta_\bfk,\rho_\bfk)}^{-1}(t)&=     -i\mqty(\rho^*_{\bfk}(t)&-\zeta_{\bfk}^*(t)\\     -\rho_{\bfk}(t)& \zeta_{\bfk}(t)).
\label{eq_alphabetat}\end{align}
Then, it is direct to write $N_{\bfk}(t)$ in terms of the free functions $W_{\bfk}(t)$, $\varphi_{\bfk}(t)$, and $Y_{\bfk}(t)$ that characterize $\zeta_{\bfk}(t)$ and $\rho_{\bfk}(t)$, and the particular solution $z_{\bfk}(t)$ defining the reference vacuum~$\ket{0}$:
\begin{align} 
    N_{\bfk}(t)&=\frac{W_{\bfk}(t)}{2}\left[1+Y_{\bfk}(t)^2\right]|z_{\bfk}(t)|^2+\frac{1}{2W_{\bfk}(t)}|\dot{z}_{\bfk}(t)|^2
    \notag\\
    &-\frac{1}{2}+Y_{\bfk}(t)\Re{z_{\bfk}^*(t)\dot{z}_{\bfk}(t)}.
\label{eq_N1}\end{align}
This first result is a generalized expression of the one found in \cite{Fedotov_2011}, which corresponds to the particular case in which we choose $(\zeta_{\bfk}(t),\rho_{\bfk}(t))$ to be the zeroth-order adiabatic approximation $(\zeta^{(0)}_{\bfk}(t),\rho^{(0)}_{\bfk}(t))$, fixed by \eqref{eq_adiabatic0} and the choice $\breve Y^{(0)}_{\bfk}(t)=0$:
\begin{equation} \label{eq_Nad}
    N^{(0)}_{\bfk}(t)=\frac{\omega_{\bfk}(t)}{2}|z_{\bfk}(t)|^2+\frac{1}{2\omega_{\bfk}(t)}|\dot{z}_{\bfk}(t)|^2-\frac{1}{2}.
\end{equation}
In this case, $z_{\bfk}(t)$ would naturally be the zeroth-order adiabatic mode (with initial adiabatic conditions at $t_0$) $\breve z^{(0)}_\bfk(t)$. In particular, our formalism also allows us to write the alternative version of this equation when we select $ {Y}^{(0)}_{\bfk}(t)=\dot{\omega}_{\bfk}(t)/[2\omega_{\bfk}(t)^2]$ instead of $\breve Y^{(0)}_{\bfk}(t)$, which provides a better zeroth-order adiabatic approximation to the equation of motion, as explained in section \ref{sec_parametrization}.

As we see from \eqref{eq_N1}, 
$N_{\bfk}(t)$ does not depend on the phase $\varphi_{\bfk}(t)$ of $\zeta_{\bfk}(t)$. Thus, although we have a freedom of three real time-dependent functions to determine the canonical quantization, the number of created particles only depends on two of them: $W_{\bfk}(t)$ and $Y_{\bfk}(t)$. This is obvious from the fact that, in our formalism where we do not ask the functions $\zeta_{\bfk}(t)$ to solve the equation of motion, multiplying $\zeta_{\bfk}(t)$ by a time-dependent phase is a trivial Bogoliubov transformation, i.e., a transformation with null $\beta$-coefficients. In the particular case that we choose $\zeta_{\bfk}(t)$ as a solution to \eqref{eq_harmonic}  related to $z_{\bfk}(t)$ by a non-trivial Bogoliubov transformation, $\dot{\varphi}_{\bfk}(t)$ would be fixed by $W_{\bfk}(t)$ according to \eqref{eq_phigammasol}. Therefore, in that case, the only freedom in the phase is its value at initial time $\varphi_{\bfk}(t_0)$, but again $N_{\bfk}(t)$ is independent from such initial value.

Once $W_{\bfk}(t)$ and $Y_{\bfk}(t)$ are chosen, in order to compute $N_{\bfk}(t)$ there is still a residual ambiguity in the choice of reference vacuum $\ket{0}$, or equivalently, in the selection of a particular solution $z_{\bfk}(t)$ to the harmonic oscillator equation \eqref{eq_harmonic} for each ${\bfk}$. However, this ambiguity can be suitably fixed  under certain circumstances. For example, let us consider matter fields which behave as in free Minkowski spacetime in the asymptotic past. This can be achieved, for instance, in the Schwinger effect by turning on the electric field at a finite time or in FLRW spacetimes by considering an asymptotically static expanding universe~\cite{birrell_davies_1982}. Then, the system possesses Poincaré symmetry when $t\rightarrow -\infty$. When we require that the quantum theory preserves this classical symmetry  in the past, we need to impose that in the asymptotic past $z_{\bfk}(t)$ behaves as a positive-frequency plane wave (according to our conventions of creation and annihilation of particles). This asymptotic condition $z_{\bfk}(t\rightarrow-\infty)$  completely determines $z_{\bfk}(t)$ for all $t$ and there remains no ambiguity in the selection of $\ket{0}$. Another example, already mentioned above,  in which there is a natural choice for $z_{\bfk}(t)$ is when the field modes behave adiabatically. In that case we are interested in comparing the $n$th-order adiabatic approximation $\left(\zeta_{\bfk}^{(n)}(t),\rho_{\bfk}^{(n)}(t)\right)$ with the corresponding exact solution, and then one chooses $z_{\bfk}(t)$ as the $n$th-order adiabatic mode~$z^{(n)}_\bfk(t)$.

In addition, there are only few cases in which it is possible to find particular solutions to~\eqref{eq_harmonic}, and hence compute $N_{\bfk}(t)$ from \eqref{eq_N1}. For example, this is the case in the Schwinger effect when the external electric field derives from a Sauter-type potential \cite{Sauter1931742}, which turns off in the asymptotic past, and following the arguments above we search for solutions that behave as positive-frequency plane waves in $t\rightarrow -\infty$ \cite{BreakingAdiabaticNavarro}. However, in general this is not possible and it would be useful to obtain a differential equation for $N_{\bfk}(t)$ in which particular solutions $z_{\bfk}(t)$ to the equations of motion do not take part explicitly. This is precisely what we are going to do in the next section.
 
\section{Generalized quantum Vlasov equation} \label{sec_GQVE}

In the following we are interested in deducing a differential equation for the number of created particles for which, unlike \eqref{eq_N1}, there is no need to solve the harmonic oscillator equation with time-dependent frequency first. Of course, this equation, just like \eqref{eq_N1}, will strongly depend on the particular choices of $(\zeta_{\bfk}(t),\rho_{\bfk}(t))$.  

The dynamics of the number of  particles as compared with the reference vacuum,  $N_{\bfk}(t)=|\beta_{\bfk}(t)|^2$, is determined by the evolution of the Bogoliubov coefficients. Hence, it will be useful to write time evolution equations for both $\alpha_{\bfk}(t)$ and $\beta_{\bfk}(t)$. For that, 
we differentiate \eqref{eq_alphabetat} with respect to $t$ and replace $\ddot{z}_{\bfk}(t)$ by $-\omega_{\bfk}(t)z_{\bfk}(t)$ as dictated by the equation of motion \eqref{eq_harmonic}. Finally, we use the inverse of \eqref{eq_alphabetat} and obtain
\begin{equation} \label{eq_evbogoliubov}
    \mqty(\dot{\alpha}_{\bfk}\\\dot{\beta}_{\bfk}^*)=i\mqty(s_{\bfk}+\dot{\varphi}_{\bfk}&r_{\bfk}e^{2i\varphi_{\bfk}}\\-r_{\bfk}^*e^{-2i\varphi_{\bfk}}&-(s_{\bfk}+\dot{\varphi}_{\bfk}))\mqty(\alpha_{\bfk}\\\beta_{\bfk}^*),
\end{equation}
where $s_{\bfk}$ is a real time-dependent function given by
\begin{equation} \label{eq_s}
    s_{\bfk}=-\frac{\omega_{\bfk}^2}{2W_{\bfk}}+\frac{1}{2}\left[\dot{Y}_{\bfk}-W_{\bfk}\left(1+Y_{\bfk}^2\right)\right]+\frac{\dot{W}_{\bfk}}{2W_{\bfk}}Y_{\bfk},
\end{equation}
while the time-dependent function $r_{\bfk}$ is determined by its real and imaginary parts, $\mu_{\bfk}$ and $\nu_{\bfk}$, respectively:
\begin{equation} \label{eq_AB}
    \mu_{\bfk}=s_{\bfk}+W_{\bfk}, \qquad\!\!
    \nu_{\bfk}=-\frac{\dot{W}_{\bfk}}{2W_{\bfk}}+W_{\bfk}Y_{\bfk}, \qquad\!\! r_{\bfk}=\mu_{\bfk}+i\nu_{\bfk}.
\end{equation}
Note that we have deliberately eliminated the dependence on the phase $\varphi_{\bfk}$ in $s_{\bfk}$ and $r_{\bfk}$, extracting it explicitly in  \eqref{eq_evbogoliubov}. Thus, both $s_{\bfk}$ and $r_{\bfk}$ are unequivocally specified once the free functions $(W_{\bfk},Y_{\bfk})$, which characterize the particular annihilation and creation operators $\hat{a}_{\bfk}(t)$ and $\hat{a}_{\bfk}(t)^{\dagger}$ in the quantum theory, are fixed. Equations \eqref{eq_evbogoliubov} coincide with the results of \cite{MottolaDeSitter99}, with the appropriate change of variables.  

Once these evolution equations are known, we generalize the procedure followed in \cite{Kluger_1998}. Differentiating $|\beta_{\bfk}(t)|^2$  and using \eqref{eq_evbogoliubov}  it can be easily seen that
\begin{equation} \label{eq_dotN}
    \dot{N}_{\bfk}(t)=2\Im{e^{-2i\varphi_{\bfk}(t)}r_{\bfk}^*(t)M_{\bfk}(t)},
\end{equation}
where we have taken advantage of the real character of $s_{\bfk}$ and   we  have defined the auxiliary function
\begin{equation}
    M_{\bfk}(t)=\alpha_{\bfk}(t)\beta_{\bfk}(t).
\end{equation}
Analogous to this deduction, it is not difficult to obtain an equation for $M_{\bfk}(t)$,
\begin{equation} \label{eq_dotM}
    \dot{M}_{\bfk}(t)=ir_{\bfk}(t)e^{2i\varphi_{\bfk}(t)}[1+2N_{\bfk}(t)]+2i[s_{\bfk}(t)+\dot{\varphi_{\bfk}}(t)]M_{\bfk}(t),
\end{equation}
by   using \eqref{eq_evbogoliubov} and the relation \eqref{eq_relationbog} between the Bogoliubov coefficients. 

Note that neither equation \eqref{eq_dotN} nor \eqref{eq_dotM} depend explicitly on the particular solution $z_{\bfk}(t)$ of the harmonic oscillator equation with time-dependent frequency. However, the residual ambiguity in the choice of reference vacuum $\ket{0}$ has not disappeared but has been transformed from the freedom in the selection of $z_{\bfk}(t)$ to the freedom in the initial conditions for $N_{\bfk}(t)$ and $M_{\bfk}(t)$.  The  natural  choice ${z}_{\bfk}(t_0)=\zeta_{\bfk}(t_0)$ and $\dot{z}_{\bfk}(t_0)=\rho_{\bfk}(t_0)$ ensures that both sets of annihilation and creation operators coincide at $t_0$, which implies 
$\beta_{\bfk}(t_0)=0$ and hence $N_{\bfk}(t_0)=M_{\bfk}(t_0)=0$.

In order to make a direct comparison with the results in the quantum kinetic approach \cite{Kluger_1998,1998,Fedotov_2011}, it will be interesting to rewrite equations \eqref{eq_dotN} and \eqref{eq_dotM} as an integro-differential equation for $N_{\bfk}(t)$ where the auxiliary function $M_{\bfk}(t)$ does not intervene. With this objective, we solve \eqref{eq_dotM} by the method of variation of constants with $N_{\bfk}$ fixed and initial condition $M_{\bfk}(t_0)=0$. Then,
\begin{equation}
    M_{\bfk}(t)=e^{2i\varphi_{\bfk}(t)}\int^t_{t_0} d\tau \  ir_{\bfk}(\tau)[1+2N_{\bfk}(\tau)]e^{i\theta_{\bfk}(t,\tau)},
\end{equation}
where 
\begin{equation} \label{eq_theta}
    \theta_{\bfk}(t,\tau)=2\int_{\tau}^t dt' \ s_{\bfk}(t').
\end{equation}
Substituting this expression in \eqref{eq_dotN} we finally obtain, in terms of the real and imaginary parts of $r_{\bfk}=\mu_{\bfk}+i\nu_{\bfk}$:
\begin{align} 
    \dot{N}_{\bfk}(t)&=\int^t_{t_0} d\tau  \ 2[1+2N_{\bfk}(\tau)]
    \notag\\
    &\times\big\{ \big[\mu_{\bfk}(t)\mu_{\bfk}(\tau)+\nu_{\bfk}(t)\nu_{\bfk}(\tau)\big]\cos[\theta_{\bfk}(t,\tau)] \nonumber \\ 
    &-\big[\mu_{\bfk}(t)\nu_{\bfk}(\tau)-\nu_{\bfk}(t)\mu_{\bfk}(\tau)\big]\sin[\theta_{\bfk}(t,\tau)]\big\}.
\label{eq_QVE2}\end{align} 
Note that $\dot{N}_{\bfk}$ does not depend on the arbitrary phase $\varphi_{\bfk}$, but only on $W_{\bfk}$ and $Y_{\bfk}$, as we already deduced in section \ref{sec_canonical}. This equation is exact and completely general for any given quantization characterized by the pair $(\zeta_{\bfk}(t),\rho_{\bfk}(t))$.

The equation above shows that pair creation is non-local in time: time evolution of $N_{\bfk}(t)$ depends on the values of this magnitude in previous times through the bosonic enhancement factor $1+2N_{\bfk}(\tau)$.\footnote{In fermionic systems, the  factor $1+2N_{\bfk}(\tau)$ transforms into a Pauli blocking factor $1-2N_{\bfk}(\tau)$ \cite{1998}.} This is due to coherence between particle creation events when intense external fields are applied. Conversely, in the limit in which external agents are weak enough, particle creation events are sufficiently separated in time so that a local approximation of this equation is feasible \cite{Kluger_1998,SchmidtNonMarkovian}.

The integro-differential equation \eqref{eq_QVE2} might seem at first sight difficult to solve. However, the canonical approach discussed in section \ref{sec_canonical} helped us to indirectly solve it. Indeed, the expression \eqref{eq_N1} for $N_{\bfk}$ is a solution to the above equation. The difficulty in solving an integro-differential equation translates into calculating a particular solution $z_{\bfk}(t)$ of the harmonic oscillator equation with time-dependent frequency \eqref{eq_harmonic}, which, as we have already discussed, can only be analytically done in specific cases such as constant external fields.

When we choose $(\zeta_{\bfk}(t),\rho_{\bfk}(t))$ as a zeroth-order adiabatic approximation \eqref{eq_adiabatic0} with $\breve Y^{(0)}_{\bfk}(t)=0$, the real time-dependent functions taking part in the previous equation reduce to
\begin{align} 
   & \breve \mu_{\bfk}^{(0)}(t)=0, \qquad \breve \nu_{\bfk}^{(0)}(t)=-\frac{\dot{\omega}_{\bfk}(t)}{2\omega_{\bfk}(t)}, \notag\\ & \breve \theta_{\bfk}^{(0)}(t,\tau)=-2\int_{\tau}^t dt' \ \omega_{\bfk}(t'),
\label{ABthetaad}\end{align}
leading to the usual integro-differential QVE found in the literature \cite{Kluger_1998}:
\begin{align}
    \dot{ N}_{\bfk}^{(0)}(t)&=\frac{\dot{\omega}_{\bfk}(t)}{2\omega_{\bfk}(t)}\int^t_{t_0} d\tau \ \frac{\dot{\omega}_{\bfk}(\tau)}{\omega_{\bfk}(\tau)}\left[1+2N_{\bfk}^{(0)}(\tau)\right]
    \notag\\
    &\times\cos\left[ 2\int^t_{\tau} dt' \ \omega_{\bfk}(t') \right].
 \label{eq_QVEadiabatic}\end{align} 
Therefore, \eqref{eq_QVE2} is the generalized QVE for arbitrary chosen functions $(\zeta_{\bfk}(t),\rho_{\bfk}(t))$. In particular, this generalization allows us to write the QVE corresponding to the zeroth-order adiabatic approximation, but with the selection $ {Y}^{(0)}_{\bfk}(t)=\dot{\omega}_{\bfk}(t)/[2\omega_{\bfk}(t)^2]$. Indeed, it is easy to verify that this equation is characterized by the functions
\begin{align} 
     {\mu}_{\bfk}^{(0)}(t)&=\frac{1}{4}\left[ \frac{\ddot{\omega}_{\bfk}(t)}{\omega_{\bfk}(t)^2}-\frac{3}{2}\frac{\dot{\omega}_{\bfk}(t)^2}{\omega_{\bfk}(t)^3} \right], \qquad  {\nu}^{(0)}_{\bfk}(t)=0, 
     \notag\\
       {\theta}_{\bfk}^{(0)}(t)&=-2\int_{\tau}^{t} dt' \ W_{\bfk}^{(2)}(t').
\label{eq_QVEYtilde}\end{align}
As explained in section \ref{sec_parametrization}, with this last choice one ensures a better adiabatic approximation to the equation of motion while maintaining the same expression for $\zeta^{(0)}_{\bfk}(t)$. Moreover, remember that this is the usual definition for the zeroth-order adiabatic mode in the context of quantum field theory in curved spacetime \cite{birrell_davies_1982}. In addition, while the only non-vanishing contribution to the usual QVE \eqref{eq_QVEadiabatic}, $\breve{\nu}_{\bfk}^{(0)}(t)$, is of first-adiabatic order, for the generalized QVE characterized by \eqref{eq_QVEYtilde} the only term which contributes, $\mu_{\bfk}^{(0)}(t)$, is of second-adiabatic order. This translates into $\dot{N}^{(0)}_{\bfk}(t)$ being of two higher adiabatic orders  for the choice of $ {Y}_{\bfk}^{(0)}(t)$ than for $\breve Y_{\bfk}^{(0)}(t)$. Thus, the generalized QVE for the choice of $ {Y}^{(0)}_{\bfk}(t)$ provides a good balance between precision and simplicity when compared to the usual QVE \eqref{eq_QVEadiabatic}.

In section \ref{sec_unitary} we will particularize the generalized QVE to quantizations that allow for a unitary implementation of the dynamics in the quantum theory, studying their relation in the ultraviolet limit with the ones for adiabatic modes.

Finally, we note that in order to perform explicit calculations it is more convenient to rewrite the integro-differential equation \eqref{eq_QVE2}, whose numerical resolution is not generally easy \cite{SchmidtNonMarkovian}, as a real linear system of ordinary differential equations.  This was first done in~\cite{Bloch1999} for the standard QVE.  To that end, we define two auxiliary time-dependent functions:
\begin{align}
    M_{1{\bfk}}(t)&=\int^t_{t_0} d\tau \ 2[1+2N_{\bfk}(\tau)]
    \notag\\
    &\times\left\{ \mu_{\bfk}(\tau)\cos[\theta_{\bfk}(t,\tau)]-\nu_{\bfk}(\tau)\sin[\theta_{\bfk}(t,\tau)] \right\}, \nonumber\\
    M_{2{\bfk}}(t)&=\int^t_{t_0} d\tau \ 2[1+2N_{\bfk}(\tau)]
    \notag\\
        &\times
        \left\{ \mu_{\bfk}(\tau)\sin[\theta_{\bfk}(t,\tau)]+\nu_{\bfk}(\tau)\cos[\theta_{\bfk}(t,\tau)] \right\},
\end{align}
such that
\begin{equation}
    \dot{N}_{\bfk}(t)=\mu_{\bfk}(t)M_{1{\bfk}}(t)+\nu_{\bfk}(t)M_{2{\bfk}}(t).
\end{equation}
Differentiating these auxiliary functions we obtain the linear differential system: 
\begin{equation} \label{eq_difsyst}
    \frac{d}{dt} \mqty(1+2N_{\bfk}\\M_{1{\bfk}}\\M_{2{\bfk}})=2\mqty(0&\mu_{\bfk}&\nu_{\bfk}\\\mu_{\bfk}&0&-s_{\bfk}\\\nu_{\bfk}&s_{\bfk}&0) \mqty(1+2N_{\bfk}\\M_{1\bfk}\\M_{2\bfk}).
\end{equation}
These real differential equations are also equivalent to the complex differential system composed by \eqref{eq_dotN} and \eqref{eq_dotM}. We have verified that this system of equations is equivalent to the one derived in \cite{MottolaDeSitter99}, that carries out an analog analysis focusing on adiabatic modes of arbitrary order.

\section{Unitary dynamics}
\label{sec_unitary}

Hamiltonians associated with the type of systems studied in this work are time-dependent. Thus, time translational invariance is broken. One usually wants the associated Fock quantum theory to preserve the symmetries of the classical system. When it no longer possesses Poincaré symmetry, as it is the case, this requirement is not restrictive enough to select particular functions $(\zeta_{\bfk}(t),\rho_{\bfk}(t))$ in general  and ambiguities emerge in the canonical quantization (see section \ref{sec_canonical}). In order to reduce these ambiguities, previous studies of both scalar and fermionic fields in homogeneous cosmological settings \cite{universe7080299,CORTEZ201536,Cortez:2019orm,Cortez:2020rla} as well as in the context of the Schwinger effect \cite{Garay2020,AlvarezFermions} impose that the canonical time evolution of the fields be unitarily implemented in the quantum theory. Physically, this translates into a well-defined total number of created particles throughout the evolution of fields at all finite times. This physical condition imposes a restriction on the large wave vector $\bfk$-functions $\zeta_{\bfk}(t)$ and $\rho_{\bfk}(t)$. The main consequence of demanding a unitary implementation of the quantum field dynamics, as proven in references \cite{universe7080299,CORTEZ201536,Cortez:2019orm,Cortez:2020rla,Garay2020,AlvarezFermions}, is its uniqueness: quantizations compatible with this requirement form a unique unitarily equivalent family. One of the primary objectives in this section is to emphasize and generalize the procedures from references \cite{CORTEZ201536,universe7080299,Cortez:2019orm,Cortez:2020rla,Garay2020,AlvarezFermions} in order to extract relevant physical properties of the generalized QVE \eqref{eq_QVE2} when particularized to this unique family of quantizations. More precisely, once $\zeta_{\bfk}(t)$ and $\rho_{\bfk}(t)$ allowing for a unitary implementation of the dynamics are characterized in terms of their asymptotic ultraviolet behavior, we will see that the usual QVE \eqref{eq_QVEadiabatic} is in most cases (but not   all) the leading order of the generalized QVE~\eqref{eq_QVE2}. This analysis will motivate a new criterion to further reduce the quantization ambiguities.

\subsection{Time evolution} \label{sec_timeevolution}

First, we study time evolution as a classical Bogoliubov transformation. In addition, we are interested in comparing formalisms used in works about unitary implementation of the quantum dynamics \cite{universe7080299,Cortez:2019orm,CORTEZ201536,Cortez:2020rla,Garay2020,AlvarezFermions} and others dealing with the quantum kinetic approach \cite{Kluger_1998,MottolaDeSitter99,1998,Fedotov_2011} for deducing the usual QVE \eqref{eq_QVEadiabatic}. Moreover, this will help to simplify proofs in future sections.

Let us consider the canonical time evolution $\mathcal{T}(t_0,t)$ of the canonically conjugate fields $(\phi(t,\textbf{x}),\pi(t,\textbf{x}))$ from $t_0$ to time $t$. The pairs of modes $(\phi_{\bfk}(t),\pi_{\bfk}(t))$ are dynamically decoupled for different $\bfk$, i.e., $\phi_{\bfk}(t)$ satisfy decoupled harmonic oscillator equations \eqref{eq_harmonic}. Thus, we can write
\begin{equation} \label{eq_defTt0t}
    \mqty(\phi_{\bfk}(t)\\\pi_{\bfk}(t))=\mathcal{T}_{\bfk}(t_0,t) \mqty(\phi_{\bfk}(t_0)\\\pi_{\bfk}(t_0)),
\end{equation}
where $\mathcal{T}_{\bfk}(t_0,t)$ is the component of $\mathcal{T}(t_0,t)$ relating the $\bfk$-modes. As the annihilation and creation variables $a_{\bfk}$ and $a_{\bfk}^*$ are time-independent, from \eqref{eq_matrixz} we deduce that
\begin{equation} \label{eq_Tt0tG}
    \mathcal{T}_{\bfk}(t_0,t)=G_{(z_{\bfk},\dot{z}_{\bfk})}(t)G^{-1}_{(z_{\bfk},\dot{z}_{\bfk})}(t_0).
\end{equation}
Note that, although the fundamental matrix $G_{(z_{\bfk},\dot{z}_{\bfk})}(t)$ depends on the particular solution $z_{\bfk}(t)$ to the equation of motion \eqref{eq_harmonic} that we had chosen, according to the general knowledge about linear ordinary differential equations, the canonical matrix $\mathcal{T}_{\bfk}(t_0,t)$ is independent of~$z_{\bfk}(t)$.

The  time evolution transformation $\mathcal{T}(t_0,t)$ has an associated Bogoliubov transformation $\mathcal{\tilde{B}}(t_0,t)$   whose  $\bfk$-component $\mathcal{\tilde{B}}_{\bfk}(t_0,t)$ relates the initial conditions $a_{\bfk}(t_0)$ and $a_{\bfk}^*(t_0)$ for the creation and annihilation variables to their time-evolved ones, i.e.,
\begin{align} 
    \mqty(a_{\bfk}(t)\\a_{\bfk}^*(t))&=\mathcal{\tilde{B}}_{\bfk}(t_0,t) \mqty(a_{\bfk}(t_0)\\a_{\bfk}^*(t_0)), 
    \notag\\
     \mathcal{\tilde{B}}_{\bfk}(t_0,t)&=\mqty(\tilde{\alpha}_{\bfk}(t_0,t)&\tilde{\beta}_{\bfk}(t_0,t)\\\tilde{\beta}_{\bfk}^*(t_0,t)&\tilde{\alpha}_{\bfk}^*(t_0,t)).
\label{eq_Bogoliubovtime}\end{align}
Therefore, at the quantum level, $\mathcal{\tilde{B}}_{\bfk}(t_0,t)$ compares the quantum theories defined by the same choice of   $(\zeta_{\bfk}(t),\rho_{\bfk}(t))$  at two different times $t_0$ and $t$, characterized by annihilation and creation operators $(\hat{a}_{\bfk}(t_0),\hat{a}_{\bfk}(t_0)^{\dagger})$ and $(\hat{a}_{\bfk}(t),\hat{a}_{\bfk}(t)^{\dagger})$, respectively. 

Explicitly, the relation between the time-evolution transformation $\mathcal{T}_{\bfk}(t_0,t)$  defined in~\eqref{eq_defTt0t} and $\mathcal{\tilde{B}}_{\bfk}(t_0,t)$ reads
\begin{equation} \label{eq_Bt0tG}
    \mathcal{\tilde{B}}_{\bfk}(t_0,t)=G^{-1}_{(\zeta_{\bfk},\rho_{\bfk})}(t)\mathcal{T}_{\bfk}(t_0,t)G_{(\zeta_{\bfk},\rho_{\bfk})}(t_0),
\end{equation}
as it is easy to deduce using  \eqref{eq_matrix}. Written in this way it is clear how  the time-dependent transformations $G_{(\zeta_{\bfk},\rho_{\bfk})}(t)$ for each $\bfk$ mediate between the classical time-evolution of the field $\mathcal{T}(t_0,t)=\oplus_{\bfk} \mathcal{T}_{\bfk}(t_0,t)$ and the Bogoliubov transformation $\mathcal{\tilde{B}}(t_0,t)=\oplus_{\bfk} \mathcal{\tilde{B}}_{\bfk}(t_0,t)$ that relates Fock quantizations at different times. It is the latter that encodes the quantum field dynamics and therefore the transformation that one would like to implement via a unitary operator $\hat{U}(t_0,t)=\oplus_{\bfk}\hat{U}_{\bfk}(t_0,t)$ such that
\begin{equation} \label{eq_unitary}
    \mqty(\hat{a}_{\bfk}(t)\\\hat{a}_{\bfk}(t)^{\dagger})=\hat{U}_{\bfk}(t_0,t) \mqty(\hat{a}_{\bfk}(t_0)\\\hat{a}_{\bfk}(t_0)^{\dagger})\hat{U}_{\bfk}(t_0,t)^{-1} .
\end{equation}
This is a non-trivial question, and only appropriate choices of $ (\zeta_{\bfk}(t),\rho_{\bfk}(t))$ render $\mathcal{\tilde{B}}(t_0,t)$ unitarily implementable at the quantum level \cite{CORTEZ201536}, as we will discuss later.

Note the difference between this time evolution Bogoliubov transformation $\mathcal{\tilde{B}}(t_0,t)$ and the previously considered $\mathcal{B}(t)$, defined by \eqref{eq_bogoliubov}. $\mathcal{B}_{\bfk}(t)$ relates the reference Fock quantization associated with a particular solution $z_{\bfk}(t)$ of the harmonic oscillator equation~\eqref{eq_harmonic} (with annihilation and creation variables denoted by $a_{\bfk}$ and $a_{\bfk}^*$) to another canonical quantization defined by chosen functions $(\zeta_{\bfk}(t),\rho_{\bfk}(t))$ (associated with $a_{\bfk}(t)$ and $a_{\bfk}^*(t)$).  While $\mathcal{B}(t)$ is usually studied in works concerning the quantum kinetic approach and the QVE \cite{Kluger_1998,MottolaDeSitter99,1998,Fedotov_2011}, the works that study the uniqueness of the quantizations that unitarily implement the dynamics, such as \cite{universe7080299,CORTEZ201536, Cortez:2019orm,Cortez:2020rla,Garay2020,AlvarezFermions}, deal with the time evolution Bogoliubov transformation $\mathcal{\tilde{B}}(t_0,t)$.

For the present study, it is useful to find a relation between $\mathcal{\tilde{B}}(t_0,t)$ and $\mathcal{B}(t)$.  The latter can be computed from \eqref{eq_alphabetat}, and 
using \eqref{eq_Tt0tG} we can rewrite \eqref{eq_Bt0tG} as
\begin{equation} \label{eq_relBt0tBt}
    \mathcal{\tilde{B}}_{\bfk}(t_0,t)=\mathcal{B}_{\bfk}(t)\mathcal{B}_{\bfk}(t_0)^{-1}.
\end{equation}
Note that this decomposition explicitly depends on the reference quantization. We can interpret the Bogoliubov transformation $\mathcal{\tilde{B}}_{\bfk}(t_0,t)$ implementing the time evolution of the field as a composition of two canonical transformations. First, $\mathcal{B}_{\bfk}(t_0)^{-1}$ transforms the initial conditions $(a_{\bfk}(t_0),a_{\bfk}^*(t_0))$ into the time-independent annihilation and creation variables $(a_{\bfk},a_{\bfk}^*)$ associated with the particular solution $z_{\bfk}(t)$. Second, $\mathcal{B}_{\bfk}(t)$ takes $(a_{\bfk},a_{\bfk}^*)$ to the time-evolved $(a_{\bfk}(t),a_{\bfk}^*(t))$. In other words, by means of an auxiliary set of modes $z_{\bfk}(t)$ we have factorized $\mathcal{\tilde{B}}_{\bfk}(t_0,t)$ in terms of Bogoliubov transformations relating $\zeta_{\bfk}(t)$ and $z_{\bfk}(t)$ at different times.

Given $\mathcal{\tilde{B}}_{\bfk}(t_0,t)$, we can define the magnitude $\tilde{N}_{\bfk}(t_0,t)=|\tilde{\beta}_{\bfk}(t_0,t)|^2$. It measures the number of created particles at some instant $t$ from the evolution of the vacuum defined at $t_0$, which is the state annihilated by all the operators $\hat{a}_{\bfk}(t_0)$. As discussed previously, we will naturally choose $z_{\bfk}(t_0)=\zeta_{\bfk}(t_0)$, and hence $\dot{z}_{\bfk}(t_0)=\rho_{\bfk}(t_0)$; equivalently, $\hat{a}_{\bfk}(t_0)=\hat{a}_{\bfk}$. Then, that vacuum state is just the reference vacuum $|0\rangle$ of previous sections, and we would simply obtain $\mathcal{\tilde{B}}_{\bfk}(t_0,t)=\mathcal{B}_{\bfk}(t)$. Then, both notions of the number of created particles coincide: $\tilde{N}_{\bfk}(t_0,t)=N_{\bfk}(t)=|\beta_{\bfk}(t)|^2$,
with $N_{\bfk}(t_0)=0$.  This will simplify the study of the unitary implementation of the dynamics in the next section.

\subsection{Unitary implementation of the dynamics}

In this section we will characterize those quantizations that unitarily implement the quantum field dynamics. For concreteness, we will restrict our arguments to the scalar Schwinger effect, already studied in \cite{Garay2020}. Here we will review the results that we need for our analysis, also adapting them to our present formalism. Other references \cite{Cortez:2019orm,Cortez:2020rla,CORTEZ201536,universe7080299} have already studied this in cosmological settings and in the fermionic Schwinger effect~\cite{AlvarezFermions}. This section depends on the particularities of the system, as we will use the asymptotic dependence of frequencies on its label (wave number $\bfk$ in the Schwinger effect, for example). Instead of working with $Y_{\bfk}(t)$ fixed to zero as it is often done, we will also study the restrictions imposed on it. This is interesting due to the fact that $N_{\bfk}(t)$ depends on it (see \eqref{eq_N1} and~\eqref{eq_QVE2}). 

A theorem by Shale \cite{Shale:1962,RUIJSENAARS1978105} ensures that $\mathcal{\tilde{B}}(t_0,t)$ is unitarily implementable if and only if the total number of created particles in the evolution of the field,
\begin{equation} \label{eq_Shale}
    \int d^3\bfk  \tilde{N}_{\bfk}(t_0,t)=\int_0^{2\pi}d\phi \int_0^{\pi}d\theta  \sin{\theta} \int_0^{\infty}dk  k^2\tilde{N}_{\bfk}(t_0,t),
\end{equation}
is finite for each fixed finite time $t$.\footnote{For other systems the integral might be substituted by a sum over the discrete indexes enumerating the frequencies, with their corresponding degeneracies; e.g., a sum in $n$ (see \eqref{eq_omegan}) for closed FLRW spacetimes with spherical spatial symmetry.} Note that the notion of unitary implementation of the Bogoliubov transformation $\mathcal{\tilde{B}}(t_0,t)$ involves all its $\bfk$-components $\mathcal{\tilde{B}}_{\bfk}(t_0,t)$. Since $\tilde{N}_{\bfk}(t_0,t)=N_{\bfk}(t)=|\beta_{\bfk}(t)|^2$,  the unitary implementation of $\mathcal{\tilde{B}}(t_0,t)$ is satisfied if and only if in the ultraviolet limit $|\bfk|=k\rightarrow \infty$ we have
\begin{equation} \label{eq_unitarycondition}
    \beta_{\bfk}(t)=\order{k^{-\lambda}},\qquad \lambda>3/2,
\end{equation}
at all finite times $t$ and for almost all fixed directions $(\theta,\phi)$.\footnote{Note that we only consider ultraviolet divergences because we deal with massive scalar fields and, consequently, there are no infrared divergences.} Note that because of the anisotropy of the Schwinger effect, the ultraviolet behavior of $\beta_{\bfk}(t)$ depends on the direction in which we calculate the limit of large $k$. Indeed, from \eqref{eq_omegaSchwinger} we see that the time derivatives of the frequencies carry a leading order contribution $\dot{\omega}_{\bfk}(t)=\order{k^0}$ for directions with constant $\theta\in (0,\pi)$, while in the direction parallel to the vector potential ($\theta=0,\pi$), $\dot{\omega}_{\bfk}(t)=\order{k^{-1}}$. However, this axis has zero measure in $\mathbb{R}^3$ and does not contribute to the integral in~\eqref{eq_Shale}.

Remember that $\beta_{\bfk}(t)$ depends both on the  particular reference solution $z_{\bfk}(t)$ of the harmonic oscillator equation \eqref{eq_harmonic} and the functions $(\zeta_{\bfk}(t),\rho_{\bfk}(t))$. As we said before, in the most realistic case in which the electric field is switched off in the asymptotic past, there is no ambiguity in the selection of $z_{\bfk}(t)$ as at that initial time we are forced to choose positive-frequency plane waves for all $\bfk$. Furthermore, assuming general mild conditions on the time dependence of the frequencies\footnote{In the scalar Schwinger effect, a sufficient condition to satisfy this mild condition is that $\dot{\omega}_{\bfk}(t)/\omega_{\bfk}(t)$ both remains finite and changes it signs a finite number of times in each closed interval of time.}, reference \cite{Garay2020} proves that this particular solution behaves in the ultraviolet as
\begin{equation} \label{eq_zUV}
    |z_{\bfk}(t)|^2=\order{k^{-1}}, \quad \dot{z}_{\bfk}(t)=i\left[-\omega_{\bfk}(t)+\Lambda_{\bfk}(t)\right]z_{\bfk}(t),
\end{equation}
where $\Lambda_{\bfk}(t)$ converges to zero at least as fast as $\order{k^{-1}}$. Once $z_{\bfk}(t)$ is fixed, let us characterize the functions $(\zeta_{\bfk}(t),\rho_{\bfk}(t))$ which verify the unitary dynamics condition \eqref{eq_unitarycondition}. Using \eqref{eq_alphabetat} and \eqref{eq_zUV}, we can write $\beta_{\bfk}(t)$ as
\begin{align} 
    \beta_{\bfk}(t)&=\bigg\{ \sqrt{\frac{W_{\bfk}(t)}{2}}[1+iY_{\bfk}(t)]
    \notag\\
    &+\frac{1}{\sqrt{2W_{\bfk}(t)}}[-\omega_{\bfk}(t)+\Lambda^*_{\bfk}(t)] \bigg\} e^{i\varphi_{\bfk}(t)}z^*_{\bfk}(t).
\label{eq_betaLO}\end{align}
We see that both its real and its imaginary parts are $\order{k^{-\lambda}}$ if and only if $W_{\bfk}(t)$ and $Y_{\bfk}(t)$ behave in the ultraviolet as
\begin{equation} \label{eq_unitarysigma}
    W_{\bfk}(t)=\omega_{\bfk}(t)\big[1+\order{k^{-\gamma}}\big],   \qquad Y_{\bfk}(t)=\order{k^{-\eta}}, 
\end{equation}
with $\gamma,\eta>3/2$,
for each finite time $t$ and for almost all $\bfk$. These two conditions characterize the choice of $(\zeta_{\bfk}(t),\rho_{\bfk}(t))$ that allow for a unitary implementation of the dynamics. 

All adiabatic approximations of arbitrary order are in this family as they all behave in the ultraviolet as $W^{(n)}_{\bfk}(t)=\omega_{\bfk}(t)\big[1+\order{k^{-3}}\big]$, $Y^{(n)}_{\bfk}(t)=\order{k^{-2}}$, and $\breve{Y}^{(n)}_{\bfk}(t)=\order{k^{-2}}$. However, without any additional criteria we cannot distinguish them from the rest of the possible choices that allow for a unitary implementation of the dynamics. On the other hand, note that as long as the external agent is time-dependent, the usual Minkowski positive-frequency plane wave modes 
do not allow for a unitary implementation of the dynamics since, for this quantization, $\gamma=1$ (see e.g. references  ~\cite{Nonadiabatic_Kim,Huet_2014} for the corresponding QVE).  Then, using Minkowski modes in the Schwinger effect would lead to  finite values of $N_{\bfk}(t)$ when the electric field is turned on  but the sum of all of them would diverge~\cite{Ruijsencharged}. 

As an aside, in cosmological isotropic settings such as in FLRW spacetimes, the behavior of the time derivative of the frequencies does not depend on the angle $\theta$. For instance, frequencies would be of the form $\omega_{\bfk}(t)~=~\sqrt{k^2+m(t)^2}$, where $m(t)$ is independent of $\bfk$, and hence $\dot{\omega}_{\bfk}(t)=\order{k^{-1}}$ in all directions. An analogous analysis to the one developed here leads to the same behavior of functions \eqref{eq_unitarysigma}.

In summary, in order to completely fix the canonical quantization scheme, we started with a freedom of three real time-dependent functions $(W_{\bfk}(t),Y_{\bfk}(t),\varphi_{\bfk}(t))$ for each $\bfk$. We proved in sections \ref{sec_N(t)} and \ref{sec_GQVE} that the number of created particles $N_{\bfk}(t)$ does not depend on $\varphi_{\bfk}(t)$. Now, the behavior of functions $W_{\bfk}(t)$ and $Y_{\bfk}(t)$ with large $k=|\bfk|$ have been restricted. Moreover, reference \cite{Garay2020}
shows that the possible selections compatible with these restrictions form a unique unitarily equivalent family of Fock quantizations in which the total number of created particles (sum of all the contributions by each~$\bfk$) remains finite at all finite times. Nevertheless, it is important to remember that each particular selection in the family provides a different total number of particles.

\subsection{Generalized QVE and unitary quantum dynamics}

In the following we are going to study the asymptotic ultraviolet behavior of the generalized QVE \eqref{eq_QVE2} for canonical quantizations unitarily implementing the dynamics. In particular, this study will provide us with an additional physical criterion, stronger than the unitary implementation of the dynamics, to reduce the ambiguities in the canonical quantization.

In the ultraviolet, our system should resemble free Minkowski spacetime regardless of the curvature or the external fields at work. This suggests a kind of generic ultraviolet behavior for the generalized QVE, independent of the specifics of the canonical quantization, at leading order. Such details should certainly play a role in subleading terms. 

Actually, we are going to argue that quantizations that unitarily implement the dynamics, i.e., with the ultraviolet behavior \eqref{eq_unitarysigma}, satisfy this criterion provided that $\gamma,\eta>2$. Otherwise the leading order of the generalized QVE depends on the specific quantization that is being carried out. It turns out that this generic quantization-independent leading order is precisely that of the QVE, i.e., the  generalized QVE for zeroth-order adiabatic modes.

Indeed, let us consider a canonical quantization defined by functions $W_{\bfk}(t)$ and $Y_{\bfk}(t)$ which behave  in the ultraviolet according to the unitary dynamics requirement \eqref{eq_unitarysigma} but with the stronger condition
\begin{equation} \label{eq_newcriterion}
    W_{\bfk}(t)=\omega_{\bfk}(t)\big[1+\order{k^{-\gamma}}\big],   \qquad Y_{\bfk}(t)=\order{k^{-\eta}},
\end{equation}
with $\gamma,\eta>2$.
This faster ultraviolet decay implies  that the leading order of its generalized QVE \eqref{eq_QVE2} coincides with the usual QVE \eqref{eq_QVEadiabatic} as can be seen by straightforward calculation. Thus, for generic functions $W_{\bfk}(t)$ and $Y_{\bfk}(t)$ in this subfamily of quantizations allowing for a unitary implementation of the dynamics, the ultraviolet behavior of $\dot{N}_{\bfk}(t)$ at leading order is independent of the particular time-dependence of those functions other than that imposed by the external electric field through $\omega_\bfk(t)$.

On the other hand, when generic canonical quantizations allow for a unitary implementation of the dynamics  but do not satisfy the previous stronger condition \eqref{eq_newcriterion}, their generalized QVE provides particle creation rates $\dot{N}_{\bfk}(t)$ whose ultraviolet behaviors at leading order strongly depend on functions $W_{\bfk}(t)$ and $Y_{\bfk}(t)$ themselves, even with a slower ultraviolet decay. More precisely, under these hypotheses the leading orders in the expansions in $k=|\bfk|$ of the functions \eqref{eq_AB} defining the generalized QVE are
\begin{align}
    \mu_{\bfk}|_{\text{L.O.}}=&2k\left.\left(1-\sqrt{\frac{\omega_{\bfk}}{W_{\bfk}}}\right)\right|_{\text{L.O.}}=\order{k^{1-\gamma}}, \nonumber\\
    \nu_{\bfk}|_{\text{L.O.}}=&-\frac{1}{2}k^{-1}q\dot{A}\cos{\theta}+kY_{\bfk}|_{\text{L.O.}}
    \notag\\ =&\order{k^{-1}}+\order{k^{1-\eta}}.
\end{align}
When $\gamma<2$ or $\eta<2$, one of them converges to zero more slowly than $\breve{\nu}_{\bfk}^{(0)}(t)=\order{k^{-1}}$ in the case of the usual QVE for the lowest adiabatic approximation (see~\eqref{ABthetaad}).  
The limiting  cases $\gamma=2$ and $\eta\geq 2$ and vice versa lead
to the same ultraviolet decay as the generic one (for $\gamma,\eta~>~2$) but in a state-dependent fashion.

Note that this analysis is valid as long as the leading order of the generalized QVE is of the same adiabatic order as the standard QVE. There are a few exceptions to this generic case such as   canonical quantizations based on higher-order adiabatic approximations, whose generalized QVE are of higher order. The $n$th-adiabatic approximation cancels the lower order contributions to its generalized QVE, and in particular, that of the usual QVE. But this makes the leading order being that of the $n$th-adiabatic approximation, which decays faster than the zeroth-order in the ultraviolet, significantly diminishing the rate of particle creation. More explicitly, the leading order of the functions $\mu^{(n)}_{\bfk}$ and $\nu^{(n)}_{\bfk}$ for the $n$th-order adiabatic modes (with $n\geq 2$) are
\begin{align}
    \mu^{(n)}_{\bfk}|_{\text{L.O.}}&=W_{\bfk}^{(n)}-W_{\bfk}^{(n+2)}=\mathcal O (k^{-(n+2)}), \nonumber\\
    \nu^{(n)}_{\bfk}|_{\text{L.O.}}&=k\big(Y_{\bfk}^{(n)}-Y_{\bfk}^{(n+2)}\big)=\mathcal O (k^{-(n+3)}).
\end{align}

For all these reasons, we consider that the physically reasonable choices for generic $W_{\bfk}(t)$ and $Y_{\bfk}(t)$ should satisfy \eqref{eq_newcriterion}. This allows not only for a unitary implementation of the dynamics, but also provides a generalized QVE whose leading order coincides with the corresponding to the lowest adiabatic order.   Other selections not satisfying this criterion but such that they cancel the contribution for the usual QVE (e.g., higher order-adiabatic approximations, which have $Y^{(n)}_{\bfk}=\order{k^{-2}}$), lead to particle creation rates which converge even faster to zero than all the others as we have discussed and are therefore good candidates as well.

 One could also consider more restricting criteria in the task of reducing the ambiguity in the quantization, based on the generalized QVE for higher adiabatic orders. A motivation for these criteria may come from the fact that, in cosmological settings and within the strict family of adiabatic vacua, it is necessary to consider higher adiabatic orders to obtain a well-defined renormalized stress-energy tensor \cite{PhysRevD.9.341}. 

\section{Conclusions} 
\label{sec_conclusions}

In usual quantum field theory in Minkowski spacetime, Poincaré symmetry fixes the vacuum. In curved spacetimes, or if an external agent is coupled to a matter field in flat spacetime, the classical system is, in general, not invariant under such a restrictive group of symmetry. In particular, when time translational invariance is lost, the vacuum changes throughout time and particle creation effects can occur. When imposing that the associate quantum theory preserves the classical symmetries we find that there are still ambiguities in the choice of vacuum defining the quantum theory.

In this work we have written a generalized version of the usual quantum Vlasov equation~\cite{Kluger_1998}, which is an integro-differential equation for the number of created particles throughout time for the Schwinger effect, extending it to arbitrary canonical quantizations. We have also provided its formal solution, thus generalizing the result in \cite{Fedotov_2011}. Moreover, we have particularized it for arbitrary $n$th-order adiabatic modes, calculating its leading order in an adiabatic expansion.

Although our analysis has been carried out for the scalar Schwinger effect, in which an external homogeneous time-dependent electric field is applied in flat spacetime, we  have also argued how our analysis can be straightforwardly applied to quantum matter fields propagating in FLRW spacetimes.

Next, we have resorted to the unitary implementation of the quantum field dynamics as physical criterion to restrict the set of acceptable quantizations. This criterion, mainly pushed forward in the context of FLRW cosmological spacetimes \cite{Cortez:2019orm,Cortez:2020rla,universe7080299,CORTEZ201536}, reduces the ambiguities in the canonical quantization to a unique family of unitarily equivalent quantizations. This also happens to be true in the scalar and fermionic Schwinger effects with a homogeneous electric field \cite{Garay2020,AlvarezFermions}. In practice, this requirement restricts the ultraviolet behavior of the Fourier modes used in the quantization so that the total number of created particles is well-defined at all finite times. 

Focusing on the quantizations that allow for a unitary implementation of the dynamics, in the present work we have proved that there is a wide family of them whose generalized QVE behaves, at leading order in the ultraviolet asymptotic expansion, exactly as the standard QVE for zeroth-order adiabatic modes. Namely, the time dependence of such leading order is only due to the characteristics of the external agent (electric field) responsible for the creation of particles, and not to the specific modes used to quantize our field.
On the other hand, we have also proved that there is another family of quantizations that, while also allowing for a unitary implementation of the dynamics, yield a generalized QVE whose leading order in the ultraviolet limit depends explicitly on the quantization (via a time dependent term that is not simply determined by the time dependence of the external agent). In view of this last result we have proposed a new criterion which, together with the unitary implementation of the dynamics, restricts even more the quantizations that we consider acceptable: those for which the leading order of the generalized QVE is just that of the zeroth-order adiabatic vacuum  (except when this leading order vanishes, e.g. for the higher order adiabatic vacua). This criterion guarantees that the particle creation rate is independent of the details of the quantization at leading order in the ultraviolet, and which decays at least as fast as for the lowest adiabatic approximation.

\acknowledgments

This work has been supported by Project. No. MICINN PID2020-118159GB-C44 from Spain.
AAD acknowledges financial support from Universidad Complutense de Madrid through the predoctoral Grant No. CT82-~20.

\bibliography{QVE}

\providecommand{\noopsort}[1]{}\providecommand{\singleletter}[1]{#1}%
\begin{thebibliography}{52}%
\makeatletter
\providecommand \@ifxundefined [1]{%
 \@ifx{#1\undefined}
}%
\providecommand \@ifnum [1]{%
 \ifnum #1\expandafter \@firstoftwo
 \else \expandafter \@secondoftwo
 \fi
}%
\providecommand \@ifx [1]{%
 \ifx #1\expandafter \@firstoftwo
 \else \expandafter \@secondoftwo
 \fi
}%
\providecommand \natexlab [1]{#1}%
\providecommand \enquote  [1]{``#1''}%
\providecommand \bibnamefont  [1]{#1}%
\providecommand \bibfnamefont [1]{#1}%
\providecommand \citenamefont [1]{#1}%
\providecommand \href@noop [0]{\@secondoftwo}%
\providecommand \href [0]{\begingroup \@sanitize@url \@href}%
\providecommand \@href[1]{\@@startlink{#1}\@@href}%
\providecommand \@@href[1]{\endgroup#1\@@endlink}%
\providecommand \@sanitize@url [0]{\catcode `\\12\catcode `\$12\catcode
  `\&12\catcode `\#12\catcode `\^12\catcode `\_12\catcode `\%12\relax}%
\providecommand \@@startlink[1]{}%
\providecommand \@@endlink[0]{}%
\providecommand \url  [0]{\begingroup\@sanitize@url \@url }%
\providecommand \@url [1]{\endgroup\@href {#1}{\urlprefix }}%
\providecommand \urlprefix  [0]{URL }%
\providecommand \Eprint [0]{\href }%
\providecommand \doibase [0]{https://doi.org/}%
\providecommand \selectlanguage [0]{\@gobble}%
\providecommand \bibinfo  [0]{\@secondoftwo}%
\providecommand \bibfield  [0]{\@secondoftwo}%
\providecommand \translation [1]{[#1]}%
\providecommand \BibitemOpen [0]{}%
\providecommand \bibitemStop [0]{}%
\providecommand \bibitemNoStop [0]{.\EOS\space}%
\providecommand \EOS [0]{\spacefactor3000\relax}%
\providecommand \BibitemShut  [1]{\csname bibitem#1\endcsname}%
\let\auto@bib@innerbib\@empty
%</preamble>
\bibitem [{\citenamefont {Sauter}(1931)}]{Sauter1931742}%
  \BibitemOpen
  \bibfield  {author} {\bibinfo {author} {\bibfnamefont {F.}~\bibnamefont
  {Sauter}},\ }\bibfield  {title} {\bibinfo {title} {{\"Uber das Verhalten
  eines Elektrons im homogenen elektrischen Feld nach der relativistischen
  Theorie Diracs}},\ }\href {https://doi.org/10.1007/BF01339461} {\bibfield
  {journal} {\bibinfo  {journal} {Z. Phys.}\ }\textbf {\bibinfo {volume}
  {69}},\ \bibinfo {pages} {742} (\bibinfo {year} {1931})}\BibitemShut
  {NoStop}%
\bibitem [{\citenamefont {Schwinger}(1951)}]{Schwinger1951}%
  \BibitemOpen
  \bibfield  {author} {\bibinfo {author} {\bibfnamefont {J.~S.}\ \bibnamefont
  {Schwinger}},\ }\bibfield  {title} {\bibinfo {title} {On gauge invariance and
  vacuum polarization},\ }\href {https://doi.org/10.1103/PhysRev.82.664}
  {\bibfield  {journal} {\bibinfo  {journal} {Phys. Rev.}\ }\textbf {\bibinfo
  {volume} {82}},\ \bibinfo {pages} {664} (\bibinfo {year} {1951})}\BibitemShut
  {NoStop}%
\bibitem [{\citenamefont {Hawking}(1974)}]{BlackHexplosons}%
  \BibitemOpen
  \bibfield  {author} {\bibinfo {author} {\bibfnamefont {S.~W.}\ \bibnamefont
  {Hawking}},\ }\bibfield  {title} {\bibinfo {title} {Black hole explosions},\
  }\href {https://doi.org/10.1038/248030a0} {\bibfield  {journal} {\bibinfo
  {journal} {Nature}\ }\textbf {\bibinfo {volume} {248}},\ \bibinfo {pages}
  {30} (\bibinfo {year} {1974})}\BibitemShut {NoStop}%
\bibitem [{\citenamefont {Parker}(1969)}]{PhysRev.183.1057}%
  \BibitemOpen
  \bibfield  {author} {\bibinfo {author} {\bibfnamefont {L.}~\bibnamefont
  {Parker}},\ }\bibfield  {title} {\bibinfo {title} {{Quantized Fields and
  Particle Creation in Expanding Universes. I}},\ }\href
  {https://doi.org/10.1103/PhysRev.183.1057} {\bibfield  {journal} {\bibinfo
  {journal} {Phys. Rev.}\ }\textbf {\bibinfo {volume} {183}},\ \bibinfo {pages}
  {1057} (\bibinfo {year} {1969})}\BibitemShut {NoStop}%
\bibitem [{\citenamefont {Gibbons}\ and\ \citenamefont
  {Hawking}(1977)}]{PhysRevD.15.2738}%
  \BibitemOpen
  \bibfield  {author} {\bibinfo {author} {\bibfnamefont {G.~W.}\ \bibnamefont
  {Gibbons}}\ and\ \bibinfo {author} {\bibfnamefont {S.~W.}\ \bibnamefont
  {Hawking}},\ }\bibfield  {title} {\bibinfo {title} {{Cosmological event
  horizons, thermodynamics, and particle creation}},\ }\href
  {https://doi.org/10.1103/PhysRevD.15.2738} {\bibfield  {journal} {\bibinfo
  {journal} {Phys. Rev. D}\ }\textbf {\bibinfo {volume} {15}},\ \bibinfo
  {pages} {2738} (\bibinfo {year} {1977})}\BibitemShut {NoStop}%
\bibitem [{\citenamefont {Dabrowski}\ and\ \citenamefont
  {Dunne}(2014)}]{superadiabatic1}%
  \BibitemOpen
  \bibfield  {author} {\bibinfo {author} {\bibfnamefont {R.}~\bibnamefont
  {Dabrowski}}\ and\ \bibinfo {author} {\bibfnamefont {G.~V.}\ \bibnamefont
  {Dunne}},\ }\bibfield  {title} {\bibinfo {title} {{Superadiabatic particle
  number in Schwinger and de Sitter particle production}},\ }\href
  {https://doi.org/10.1103/physrevd.90.025021} {\bibfield  {journal} {\bibinfo
  {journal} {Phys. Rev. D}\ }\textbf {\bibinfo {volume} {90}},\ \bibinfo
  {pages} {025021} (\bibinfo {year} {2014})}\BibitemShut {NoStop}%
\bibitem [{\citenamefont {Dabrowski}\ and\ \citenamefont
  {Dunne}(2016)}]{superadiabatic2}%
  \BibitemOpen
  \bibfield  {author} {\bibinfo {author} {\bibfnamefont {R.}~\bibnamefont
  {Dabrowski}}\ and\ \bibinfo {author} {\bibfnamefont {G.~V.}\ \bibnamefont
  {Dunne}},\ }\bibfield  {title} {\bibinfo {title} {{Time dependence of
  adiabatic particle number}},\ }\href
  {https://doi.org/10.1103/physrevd.94.065005} {\bibfield  {journal} {\bibinfo
  {journal} {Phys. Rev. D}\ }\textbf {\bibinfo {volume} {94}},\ \bibinfo
  {pages} {065005} (\bibinfo {year} {2016})}\BibitemShut {NoStop}%
\bibitem [{\citenamefont {Ilderton}(2022)}]{IldertonZerothOrder}%
  \BibitemOpen
  \bibfield  {author} {\bibinfo {author} {\bibfnamefont {A.}~\bibnamefont
  {Ilderton}},\ }\bibfield  {title} {\bibinfo {title} {{Physics of adiabatic
  particle number in the Schwinger effect}},\ }\href
  {https://doi.org/10.1103/PhysRevD.105.016021} {\bibfield  {journal} {\bibinfo
   {journal} {Phys. Rev. D}\ }\textbf {\bibinfo {volume} {105}},\ \bibinfo
  {pages} {016021} (\bibinfo {year} {2022})}\BibitemShut {NoStop}%
\bibitem [{\citenamefont {Yamada}(2021)}]{Yamada_2021}%
  \BibitemOpen
  \bibfield  {author} {\bibinfo {author} {\bibfnamefont {Y.}~\bibnamefont
  {Yamada}},\ }\bibfield  {title} {\bibinfo {title} {Superadiabatic basis in
  cosmological particle production: application to preheating},\ }\href
  {https://doi.org/10.1088/1475-7516/2021/09/009} {\bibfield  {journal}
  {\bibinfo  {journal} {JCAP}\ }\textbf {\bibinfo {volume} {2021}},\ \bibinfo
  {pages} {009 (2021)}}\BibitemShut {NoStop}%
\bibitem [{\citenamefont {L^^c3^^bcders}\ and\ \citenamefont
  {Roberts}(1990)}]{LudersRoberts}%
  \BibitemOpen
  \bibfield  {author} {\bibinfo {author} {\bibfnamefont {C.}~\bibnamefont
  {L^^c3^^bcders}}\ and\ \bibinfo {author} {\bibfnamefont {J.~E.}\ \bibnamefont
  {Roberts}},\ }\bibfield  {title} {\bibinfo {title} {{Local quasiequivalence
  and adiabatic vacuum states}},\ }\href@noop {} {\bibfield  {journal}
  {\bibinfo  {journal} {Commun. Math. Phys.}\ }\textbf {\bibinfo {volume}
  {134}},\ \bibinfo {pages} {29} (\bibinfo {year} {1990})}\BibitemShut
  {NoStop}%
\bibitem [{\citenamefont {Kluger}\ \emph {et~al.}(1998)\citenamefont {Kluger},
  \citenamefont {Mottola},\ and\ \citenamefont {Eisenberg}}]{Kluger_1998}%
  \BibitemOpen
  \bibfield  {author} {\bibinfo {author} {\bibfnamefont {Y.}~\bibnamefont
  {Kluger}}, \bibinfo {author} {\bibfnamefont {E.}~\bibnamefont {Mottola}},\
  and\ \bibinfo {author} {\bibfnamefont {J.~M.}\ \bibnamefont {Eisenberg}},\
  }\bibfield  {title} {\bibinfo {title} {{Quantum Vlasov equation and its
  Markov limit}},\ }\href {https://doi.org/10.1103/physrevd.58.125015}
  {\bibfield  {journal} {\bibinfo  {journal} {Phys. Rev. D}\ }\textbf {\bibinfo
  {volume} {58}},\ \bibinfo {pages} {12} (\bibinfo {year} {1998})}\BibitemShut
  {NoStop}%
\bibitem [{\citenamefont {Habib}\ \emph {et~al.}(2000)\citenamefont {Habib},
  \citenamefont {Molina-Par^^c3^^ads},\ and\ \citenamefont
  {Mottola}}]{MottolaDeSitter99}%
  \BibitemOpen
  \bibfield  {author} {\bibinfo {author} {\bibfnamefont {S.}~\bibnamefont
  {Habib}}, \bibinfo {author} {\bibfnamefont {C.}~\bibnamefont
  {Molina-Par^^c3^^ads}},\ and\ \bibinfo {author} {\bibfnamefont
  {E.}~\bibnamefont {Mottola}},\ }\bibfield  {title} {\bibinfo {title}
  {Energy-momentum tensor of particles created in an expanding universe},\
  }\href {https://doi.org/10.1103/physrevd.61.024010} {\bibfield  {journal}
  {\bibinfo  {journal} {Phys. Rev. D}\ }\textbf {\bibinfo {volume} {61}},\
  \bibinfo {pages} {024010} (\bibinfo {year} {2000})}\BibitemShut {NoStop}%
\bibitem [{\citenamefont {Fahn}\ \emph {et~al.}(2019)\citenamefont {Fahn},
  \citenamefont {Giesel},\ and\ \citenamefont {Kobler}}]{fahn2019dynamical}%
  \BibitemOpen
  \bibfield  {author} {\bibinfo {author} {\bibfnamefont {M.~J.}\ \bibnamefont
  {Fahn}}, \bibinfo {author} {\bibfnamefont {K.}~\bibnamefont {Giesel}},\ and\
  \bibinfo {author} {\bibfnamefont {M.}~\bibnamefont {Kobler}},\ }\bibfield
  {title} {\bibinfo {title} {{Dynamical Properties of the Mukhanov-Sasaki
  Hamiltonian in the Context of Adiabatic Vacua and the Lewis-Riesenfeld
  Invariant}},\ }\href {https://doi.org/10.3390/universe5070170} {\bibfield
  {journal} {\bibinfo  {journal} {Universe}\ }\textbf {\bibinfo {volume} {5}},\
  \bibinfo {pages} {170} (\bibinfo {year} {2019})}\BibitemShut {NoStop}%
\bibitem [{\citenamefont {Elizaga~Navascu\'es}\ \emph
  {et~al.}(2019)\citenamefont {Elizaga~Navascu\'es}, \citenamefont
  {Marug\'an},\ and\ \citenamefont
  {Thiemann}}]{HamiltonianDiagonalizationGuillermo}%
  \BibitemOpen
  \bibfield  {author} {\bibinfo {author} {\bibfnamefont {B.}~\bibnamefont
  {Elizaga~Navascu\'es}}, \bibinfo {author} {\bibfnamefont {G.~A.~M.}\
  \bibnamefont {Marug\'an}},\ and\ \bibinfo {author} {\bibfnamefont
  {T.}~\bibnamefont {Thiemann}},\ }\bibfield  {title} {\bibinfo {title}
  {{Hamiltonian diagonalization in hybrid quantum cosmology}},\ }\href
  {https://doi.org/10.1088/1361-6382/ab32af} {\bibfield  {journal} {\bibinfo
  {journal} {Class. Quant. Grav.}\ }\textbf {\bibinfo {volume} {36}},\ \bibinfo
  {pages} {185010} (\bibinfo {year} {2019})}\BibitemShut {NoStop}%
\bibitem [{\citenamefont {Cortez}\ \emph
  {et~al.}(2020{\natexlab{a}})\citenamefont {Cortez}, \citenamefont
  {Elizaga~Navascu\'es}, \citenamefont {Marug\'an}, \citenamefont {Prado},\
  and\ \citenamefont {Velhinho}}]{Cortez:2020rla}%
  \BibitemOpen
  \bibfield  {author} {\bibinfo {author} {\bibfnamefont {J.}~\bibnamefont
  {Cortez}}, \bibinfo {author} {\bibfnamefont {B.}~\bibnamefont
  {Elizaga~Navascu\'es}}, \bibinfo {author} {\bibfnamefont {G.~A.~M.}\
  \bibnamefont {Marug\'an}}, \bibinfo {author} {\bibfnamefont {S.}~\bibnamefont
  {Prado}},\ and\ \bibinfo {author} {\bibfnamefont {J.~M.}\ \bibnamefont
  {Velhinho}},\ }\bibfield  {title} {\bibinfo {title} {{Uniqueness Criteria for
  the Fock Quantization of Dirac Fields and Applications in Hybrid Loop Quantum
  Cosmology}},\ }\href@noop {} {\bibfield  {journal} {\bibinfo  {journal}
  {Universe}\ }\textbf {\bibinfo {volume} {6}},\ \bibinfo {pages} {241}
  (\bibinfo {year} {2020}{\natexlab{a}})}\BibitemShut {NoStop}%
\bibitem [{\citenamefont {de~Blas}\ and\ \citenamefont
  {Olmedo}(2016)}]{PrimordialPowerJavi}%
  \BibitemOpen
  \bibfield  {author} {\bibinfo {author} {\bibfnamefont {D.~M.}\ \bibnamefont
  {de~Blas}}\ and\ \bibinfo {author} {\bibfnamefont {J.}~\bibnamefont
  {Olmedo}},\ }\bibfield  {title} {\bibinfo {title} {Primordial power spectra
  for scalar perturbations in loop quantum cosmology},\ }\href
  {https://doi.org/10.1088/1475-7516/2016/06/029} {\bibfield  {journal}
  {\bibinfo  {journal} {JCAP}\ }\textbf {\bibinfo {volume} {2016}},\ \bibinfo
  {pages} {029 (2016)}}\BibitemShut {NoStop}%
\bibitem [{\citenamefont {Elizaga~Navascu^^c3^^a9s}\ \emph
  {et~al.}(2021)\citenamefont {Elizaga~Navascu^^c3^^a9s}, \citenamefont
  {Mena~Marugan},\ and\ \citenamefont {Prado}}]{nonoscillations}%
  \BibitemOpen
  \bibfield  {author} {\bibinfo {author} {\bibfnamefont {B.}~\bibnamefont
  {Elizaga~Navascu^^c3^^a9s}}, \bibinfo {author} {\bibfnamefont {G.~A.}\
  \bibnamefont {Mena~Marugan}},\ and\ \bibinfo {author} {\bibfnamefont
  {S.}~\bibnamefont {Prado}},\ }\bibfield  {title} {\bibinfo {title}
  {{Non-oscillating power spectra in loop quantum cosmology}},\ }\href
  {https://doi.org/10.1088/1361-6382/abc6bb} {\bibfield  {journal} {\bibinfo
  {journal} {Class. Quant. Grav.}\ }\textbf {\bibinfo {volume} {38}},\ \bibinfo
  {pages} {035001} (\bibinfo {year} {2021})}\BibitemShut {NoStop}%
\bibitem [{\citenamefont {Liboff}(2003)}]{liboff}%
  \BibitemOpen
  \bibfield  {author} {\bibinfo {author} {\bibfnamefont {R.}~\bibnamefont
  {Liboff}},\ }\href@noop {} {\emph {\bibinfo {title} {{Kinetic Theory:
  Classical, Quantum, and Relativistic Descriptions}}}},\ Graduate Texts in
  Contemporary Physics\ (\bibinfo  {publisher} {Springer, New York},\ \bibinfo
  {year} {2003})\BibitemShut {NoStop}%
\bibitem [{\citenamefont {Schmidt}\ \emph {et~al.}(1998)\citenamefont
  {Schmidt}, \citenamefont {Blaschke}, \citenamefont {R^^c3^^b6pke},
  \citenamefont {Smolyansky}, \citenamefont {Prozorkevich},\ and\ \citenamefont
  {Toneev}}]{1998}%
  \BibitemOpen
  \bibfield  {author} {\bibinfo {author} {\bibfnamefont {S.}~\bibnamefont
  {Schmidt}}, \bibinfo {author} {\bibfnamefont {D.}~\bibnamefont {Blaschke}},
  \bibinfo {author} {\bibfnamefont {G.}~\bibnamefont {R^^c3^^b6pke}}, \bibinfo
  {author} {\bibfnamefont {S.~A.}\ \bibnamefont {Smolyansky}}, \bibinfo
  {author} {\bibfnamefont {A.~V.}\ \bibnamefont {Prozorkevich}},\ and\ \bibinfo
  {author} {\bibfnamefont {V.~D.}\ \bibnamefont {Toneev}},\ }\bibfield  {title}
  {\bibinfo {title} {{A Quantum Kinetic Equation for Particle Production in the
  Schwinger Mechanism}},\ }\href@noop {} {\bibfield  {journal} {\bibinfo
  {journal} {Int. J. Mod. Phys. E}\ }\textbf {\bibinfo {volume} {07}},\
  \bibinfo {pages} {709} (\bibinfo {year} {1998})}\BibitemShut {NoStop}%
\bibitem [{\citenamefont {Roberts}\ and\ \citenamefont
  {Schmidt}(2000)}]{RobertsSchmidt}%
  \BibitemOpen
  \bibfield  {author} {\bibinfo {author} {\bibfnamefont {C.~D.}\ \bibnamefont
  {Roberts}}\ and\ \bibinfo {author} {\bibfnamefont {S.~M.}\ \bibnamefont
  {Schmidt}},\ }\bibfield  {title} {\bibinfo {title} {{Dyson-Schwinger
  equations: Density, temperature and continuum strong QCD}},\ }\href
  {https://doi.org/10.1016/s0146-6410(00)90011-5} {\bibfield  {journal}
  {\bibinfo  {journal} {Prog. Part. Nucl. Phys}\ }\textbf {\bibinfo {volume}
  {45}},\ \bibinfo {pages} {S1} (\bibinfo {year} {2000})}\BibitemShut {NoStop}%
\bibitem [{\citenamefont {Ruffini}\ \emph {et~al.}(2010)\citenamefont
  {Ruffini}, \citenamefont {Vereshchagin},\ and\ \citenamefont
  {Xue}}]{Ruffini}%
  \BibitemOpen
  \bibfield  {author} {\bibinfo {author} {\bibfnamefont {R.}~\bibnamefont
  {Ruffini}}, \bibinfo {author} {\bibfnamefont {G.}~\bibnamefont
  {Vereshchagin}},\ and\ \bibinfo {author} {\bibfnamefont {S.-S.}\ \bibnamefont
  {Xue}},\ }\bibfield  {title} {\bibinfo {title} {{Electron^^e2^^80^^93positron
  pairs in physics and astrophysics: From heavy nuclei to black holes}},\
  }\href {https://doi.org/10.1016/j.physrep.2009.10.004} {\bibfield  {journal}
  {\bibinfo  {journal} {Phys. Rep.}\ }\textbf {\bibinfo {volume} {487}},\
  \bibinfo {pages} {1} (\bibinfo {year} {2010})}\BibitemShut {NoStop}%
\bibitem [{\citenamefont {Dunne}(2009)}]{Dunne}%
  \BibitemOpen
  \bibfield  {author} {\bibinfo {author} {\bibfnamefont {G.~V.}\ \bibnamefont
  {Dunne}},\ }\bibfield  {title} {\bibinfo {title} {{New strong-field QED
  effects at extreme light infrastructure}},\ }\href
  {https://doi.org/10.1140/epjd/e2009-00022-0} {\bibfield  {journal} {\bibinfo
  {journal} {Eur. Phys. J. D}\ }\textbf {\bibinfo {volume} {55}},\ \bibinfo
  {pages} {327} (\bibinfo {year} {2009})}\BibitemShut {NoStop}%
\bibitem [{\citenamefont {Dumlu}\ and\ \citenamefont
  {Dunne}(2011)}]{DumluDunne}%
  \BibitemOpen
  \bibfield  {author} {\bibinfo {author} {\bibfnamefont {C.~K.}\ \bibnamefont
  {Dumlu}}\ and\ \bibinfo {author} {\bibfnamefont {G.~V.}\ \bibnamefont
  {Dunne}},\ }\bibfield  {title} {\bibinfo {title} {{Interference effects in
  Schwinger vacuum pair production for time-dependent laser pulses}},\ }\href
  {https://doi.org/10.1103/physrevd.83.065028} {\bibfield  {journal} {\bibinfo
  {journal} {Phys. Rev. D}\ }\textbf {\bibinfo {volume} {83}},\ \bibinfo
  {pages} {065028} (\bibinfo {year} {2011})}\BibitemShut {NoStop}%
\bibitem [{\citenamefont {Hebenstreit}\ \emph {et~al.}(2009)\citenamefont
  {Hebenstreit}, \citenamefont {Alkofer}, \citenamefont {Dunne},\ and\
  \citenamefont {Gies}}]{HebenstreitShortLaserPulses}%
  \BibitemOpen
  \bibfield  {author} {\bibinfo {author} {\bibfnamefont {F.}~\bibnamefont
  {Hebenstreit}}, \bibinfo {author} {\bibfnamefont {R.}~\bibnamefont
  {Alkofer}}, \bibinfo {author} {\bibfnamefont {G.~V.}\ \bibnamefont {Dunne}},\
  and\ \bibinfo {author} {\bibfnamefont {H.}~\bibnamefont {Gies}},\ }\bibfield
  {title} {\bibinfo {title} {{Momentum Signatures for Schwinger Pair Production
  in Short Laser Pulses with a Subcycle Structure}},\ }\href
  {https://doi.org/10.1103/physrevlett.102.150404} {\bibfield  {journal}
  {\bibinfo  {journal} {Phys. Rev. Lett.}\ }\textbf {\bibinfo {volume} {102}},\
  \bibinfo {pages} {150404} (\bibinfo {year} {2009})}\BibitemShut {NoStop}%
\bibitem [{\citenamefont {Anderson}\ and\ \citenamefont
  {Mottola}(2014)}]{MottolaDeSitter}%
  \BibitemOpen
  \bibfield  {author} {\bibinfo {author} {\bibfnamefont {P.~R.}\ \bibnamefont
  {Anderson}}\ and\ \bibinfo {author} {\bibfnamefont {E.}~\bibnamefont
  {Mottola}},\ }\bibfield  {title} {\bibinfo {title} {{Instability of global de
  Sitter space to particle creation}},\ }\href
  {https://doi.org/10.1103/physrevd.89.104038} {\bibfield  {journal} {\bibinfo
  {journal} {Phys. Rev. D}\ }\textbf {\bibinfo {volume} {89}},\ \bibinfo
  {pages} {104038} (\bibinfo {year} {2014})}\BibitemShut {NoStop}%
\bibitem [{\citenamefont {Cortez}\ \emph {et~al.}(2015)\citenamefont {Cortez},
  \citenamefont {{Mena Marug\'an}},\ and\ \citenamefont
  {Velhinho}}]{CORTEZ201536}%
  \BibitemOpen
  \bibfield  {author} {\bibinfo {author} {\bibfnamefont {J.}~\bibnamefont
  {Cortez}}, \bibinfo {author} {\bibfnamefont {G.~A.}\ \bibnamefont {{Mena
  Marug\'an}}},\ and\ \bibinfo {author} {\bibfnamefont {J.~M.}\ \bibnamefont
  {Velhinho}},\ }\bibfield  {title} {\bibinfo {title} {Quantum unitary dynamics
  in cosmological spacetimes},\ }\href
  {https://doi.org/https://doi.org/10.1016/j.aop.2015.09.016} {\bibfield
  {journal} {\bibinfo  {journal} {Ann. Phys.}\ }\textbf {\bibinfo {volume}
  {363}},\ \bibinfo {pages} {36} (\bibinfo {year} {2015})}\BibitemShut
  {NoStop}%
\bibitem [{\citenamefont {Cortez}\ \emph
  {et~al.}(2020{\natexlab{b}})\citenamefont {Cortez}, \citenamefont
  {Mena~Marug\'an},\ and\ \citenamefont {Velhinho}}]{Cortez:2019orm}%
  \BibitemOpen
  \bibfield  {author} {\bibinfo {author} {\bibfnamefont {J.}~\bibnamefont
  {Cortez}}, \bibinfo {author} {\bibfnamefont {G.~A.}\ \bibnamefont
  {Mena~Marug\'an}},\ and\ \bibinfo {author} {\bibfnamefont {J.}~\bibnamefont
  {Velhinho}},\ }\bibfield  {title} {\bibinfo {title} {{Quantum Linear Scalar
  Fields with Time Dependent Potentials: Overview and Applications to
  Cosmology}},\ }\href {https://www.mdpi.com/2227-7390/8/1/115} {\bibfield
  {journal} {\bibinfo  {journal} {Mathematics}\ }\textbf {\bibinfo {volume}
  {8}},\ \bibinfo {pages} {115} (\bibinfo {year}
  {2020}{\natexlab{b}})}\BibitemShut {NoStop}%
\bibitem [{\citenamefont {Cortez}\ \emph {et~al.}(2021)\citenamefont {Cortez},
  \citenamefont {Mena~Marug^^c3^^a1n},\ and\ \citenamefont
  {Velhinho}}]{universe7080299}%
  \BibitemOpen
  \bibfield  {author} {\bibinfo {author} {\bibfnamefont {J.}~\bibnamefont
  {Cortez}}, \bibinfo {author} {\bibfnamefont {G.~A.}\ \bibnamefont
  {Mena~Marug^^c3^^a1n}},\ and\ \bibinfo {author} {\bibfnamefont {J.~M.}\
  \bibnamefont {Velhinho}},\ }\bibfield  {title} {\bibinfo {title} {{A Brief
  Overview of Results about Uniqueness of the Quantization in Cosmology}},\
  }\href {https://doi.org/10.3390/universe7080299} {\bibfield  {journal}
  {\bibinfo  {journal} {Universe}\ }\textbf {\bibinfo {volume} {7}},\ \bibinfo
  {pages} {299} (\bibinfo {year} {2021})}\BibitemShut {NoStop}%
\bibitem [{\citenamefont {Garay}\ \emph {et~al.}(2020)\citenamefont {Garay},
  \citenamefont {Mart\'in-Caro},\ and\ \citenamefont
  {Mart\'in-Benito}}]{Garay2020}%
  \BibitemOpen
  \bibfield  {author} {\bibinfo {author} {\bibfnamefont {L.~J.}\ \bibnamefont
  {Garay}}, \bibinfo {author} {\bibfnamefont {A.~G.}\ \bibnamefont
  {Mart\'in-Caro}},\ and\ \bibinfo {author} {\bibfnamefont {M.}~\bibnamefont
  {Mart\'in-Benito}},\ }\bibfield  {title} {\bibinfo {title} {{Unitary
  quantization of a scalar charged field and Schwinger effect}},\ }\href@noop
  {} {\bibfield  {journal} {\bibinfo  {journal} {JHEP}\ }\textbf {\bibinfo
  {volume} {2020}},\ \bibinfo {pages} {120 (2020)}}\BibitemShut {NoStop}%
\bibitem [{\citenamefont {\'Alvarez-Dom\'inguez}\ \emph
  {et~al.}(2021)\citenamefont {\'Alvarez-Dom\'inguez}, \citenamefont {Garay},
  \citenamefont {Garc\'ia-Heredia},\ and\ \citenamefont
  {Mart\'in-Benito}}]{AlvarezFermions}%
  \BibitemOpen
  \bibfield  {author} {\bibinfo {author} {\bibfnamefont {A.}~\bibnamefont
  {\'Alvarez-Dom\'inguez}}, \bibinfo {author} {\bibfnamefont {L.~J.}\
  \bibnamefont {Garay}}, \bibinfo {author} {\bibfnamefont {D.}~\bibnamefont
  {Garc\'ia-Heredia}},\ and\ \bibinfo {author} {\bibfnamefont {M.}~\bibnamefont
  {Mart\'in-Benito}},\ }\bibfield  {title} {\bibinfo {title} {{Quantum unitary
  dynamics of a charged fermionic field and Schwinger effect}},\ }\href
  {https://doi.org/10.1007/jhep10(2021)074} {\bibfield  {journal} {\bibinfo
  {journal} {JHEP}\ }\textbf {\bibinfo {volume} {2021}},\ \bibinfo {pages} {74
  (2021)}}\BibitemShut {NoStop}%
\bibitem [{\citenamefont {Gavrilov}\ and\ \citenamefont
  {Gitman}(1996)}]{Gavrilov:1996pz}%
  \BibitemOpen
  \bibfield  {author} {\bibinfo {author} {\bibfnamefont {S.~P.}\ \bibnamefont
  {Gavrilov}}\ and\ \bibinfo {author} {\bibfnamefont {D.~M.}\ \bibnamefont
  {Gitman}},\ }\bibfield  {title} {\bibinfo {title} {{Vacuum instability in
  external fields}},\ }\href@noop {} {\bibfield  {journal} {\bibinfo  {journal}
  {Phys. Rev. D}\ }\textbf {\bibinfo {volume} {53}},\ \bibinfo {pages} {7162}
  (\bibinfo {year} {1996})}\BibitemShut {NoStop}%
%%CITATION = HEP-TH/9603152;%%
\bibitem [{\citenamefont {Wald}(1979)}]{WALD1979490}%
  \BibitemOpen
  \bibfield  {author} {\bibinfo {author} {\bibfnamefont {R.~M.}\ \bibnamefont
  {Wald}},\ }\bibfield  {title} {\bibinfo {title} {{Existence of the S-matrix
  in quantum field theory in curved space-time}},\ }\href
  {https://doi.org/https://doi.org/10.1016/0003-4916(79)90135-0} {\bibfield
  {journal} {\bibinfo  {journal} {Ann. Phys.}\ }\textbf {\bibinfo {volume}
  {118}},\ \bibinfo {pages} {490} (\bibinfo {year} {1979})}\BibitemShut
  {NoStop}%
\bibitem [{\citenamefont {Pedrosa}(1987)}]{doi:10.1063/1.527707}%
  \BibitemOpen
  \bibfield  {author} {\bibinfo {author} {\bibfnamefont {I.~A.}\ \bibnamefont
  {Pedrosa}},\ }\bibfield  {title} {\bibinfo {title} {{Canonical
  transformations and exact invariants for dissipative systems}},\ }\href
  {https://doi.org/10.1063/1.527707} {\bibfield  {journal} {\bibinfo  {journal}
  {J. Math. Phys.}\ }\textbf {\bibinfo {volume} {28}},\ \bibinfo {pages} {2662}
  (\bibinfo {year} {1987})}\BibitemShut {NoStop}%
\bibitem [{\citenamefont {Hebenstreit}\ \emph {et~al.}(2010)\citenamefont
  {Hebenstreit}, \citenamefont {Alkofer},\ and\ \citenamefont
  {Gies}}]{Hebenstreit1}%
  \BibitemOpen
  \bibfield  {author} {\bibinfo {author} {\bibfnamefont {F.}~\bibnamefont
  {Hebenstreit}}, \bibinfo {author} {\bibfnamefont {R.}~\bibnamefont
  {Alkofer}},\ and\ \bibinfo {author} {\bibfnamefont {H.}~\bibnamefont
  {Gies}},\ }\bibfield  {title} {\bibinfo {title} {{Schwinger pair production
  in space- and time-dependent electric fields: Relating the Wigner formalism
  to quantum kinetic theory}},\ }\href
  {https://doi.org/10.1103/physrevd.82.105026} {\bibfield  {journal} {\bibinfo
  {journal} {Phys. Rev. D}\ }\textbf {\bibinfo {volume} {82}},\ \bibinfo
  {pages} {105026} (\bibinfo {year} {2010})}\BibitemShut {NoStop}%
\bibitem [{\citenamefont {Hebenstreit}\ \emph
  {et~al.}(2011{\natexlab{a}})\citenamefont {Hebenstreit}, \citenamefont
  {Ilderton}, \citenamefont {Marklund},\ and\ \citenamefont
  {Zamanian}}]{Hebenstreit2}%
  \BibitemOpen
  \bibfield  {author} {\bibinfo {author} {\bibfnamefont {F.}~\bibnamefont
  {Hebenstreit}}, \bibinfo {author} {\bibfnamefont {A.}~\bibnamefont
  {Ilderton}}, \bibinfo {author} {\bibfnamefont {M.}~\bibnamefont {Marklund}},\
  and\ \bibinfo {author} {\bibfnamefont {J.}~\bibnamefont {Zamanian}},\
  }\bibfield  {title} {\bibinfo {title} {{Strong field effects in laser pulses:
  The Wigner formalism}},\ }\href {https://doi.org/10.1103/physrevd.83.065007}
  {\bibfield  {journal} {\bibinfo  {journal} {Phys. Rev. D}\ }\textbf {\bibinfo
  {volume} {83}},\ \bibinfo {pages} {065007} (\bibinfo {year}
  {2011}{\natexlab{a}})}\BibitemShut {NoStop}%
\bibitem [{\citenamefont {Hebenstreit}\ \emph
  {et~al.}(2011{\natexlab{b}})\citenamefont {Hebenstreit}, \citenamefont
  {Ilderton},\ and\ \citenamefont {Marklund}}]{Hebenstreit3}%
  \BibitemOpen
  \bibfield  {author} {\bibinfo {author} {\bibfnamefont {F.}~\bibnamefont
  {Hebenstreit}}, \bibinfo {author} {\bibfnamefont {A.}~\bibnamefont
  {Ilderton}},\ and\ \bibinfo {author} {\bibfnamefont {M.}~\bibnamefont
  {Marklund}},\ }\bibfield  {title} {\bibinfo {title} {Pair production: The
  view from the lightfront},\ }\href
  {https://doi.org/10.1103/physrevd.84.125022} {\bibfield  {journal} {\bibinfo
  {journal} {Phys. Rev. D}\ }\textbf {\bibinfo {volume} {84}},\ \bibinfo
  {pages} {125022} (\bibinfo {year} {2011}{\natexlab{b}})}\BibitemShut
  {NoStop}%
\bibitem [{\citenamefont {Sheng}\ \emph {et~al.}(2019)\citenamefont {Sheng},
  \citenamefont {Fang}, \citenamefont {Wang},\ and\ \citenamefont
  {Rischke}}]{Schwingerwigner}%
  \BibitemOpen
  \bibfield  {author} {\bibinfo {author} {\bibfnamefont {X.}~\bibnamefont
  {Sheng}}, \bibinfo {author} {\bibfnamefont {R.}~\bibnamefont {Fang}},
  \bibinfo {author} {\bibfnamefont {Q.}~\bibnamefont {Wang}},\ and\ \bibinfo
  {author} {\bibfnamefont {D.~H.}\ \bibnamefont {Rischke}},\ }\bibfield
  {title} {\bibinfo {title} {Wigner function and pair production in parallel
  electric and magnetic fields},\ }\href
  {https://doi.org/10.1103/PhysRevD.99.056004} {\bibfield  {journal} {\bibinfo
  {journal} {Phys. Rev. D}\ }\textbf {\bibinfo {volume} {99}},\ \bibinfo
  {pages} {056004} (\bibinfo {year} {2019})}\BibitemShut {NoStop}%
\bibitem [{\citenamefont {Fonarev}(1994)}]{WignerFonarev}%
  \BibitemOpen
  \bibfield  {author} {\bibinfo {author} {\bibfnamefont {O.~A.}\ \bibnamefont
  {Fonarev}},\ }\bibfield  {title} {\bibinfo {title} {{Wigner function and
  quantum kinetic theory in curved space^^e2^^80^^93time and external
  fields}},\ }\href {https://doi.org/10.1063/1.530542} {\bibfield  {journal}
  {\bibinfo  {journal} {J. Math. Phys.}\ }\textbf {\bibinfo {volume} {35}},\
  \bibinfo {pages} {2105} (\bibinfo {year} {1994})}\BibitemShut {NoStop}%
\bibitem [{\citenamefont {Wald}(1994)}]{wald1994quantum}%
  \BibitemOpen
  \bibfield  {author} {\bibinfo {author} {\bibfnamefont {R.}~\bibnamefont
  {Wald}},\ }\href {https://books.google.es/books?id=Iud7eyDxT1AC} {\emph
  {\bibinfo {title} {{Quantum Field Theory in Curved Spacetime and Black Hole
  Thermodynamics}}}},\ Chicago Lectures in Physics\ (\bibinfo  {publisher}
  {University of Chicago Press},\ \bibinfo {year} {1994})\BibitemShut {NoStop}%
\bibitem [{\citenamefont {Birrell}\ and\ \citenamefont
  {Davies}(1982)}]{birrell_davies_1982}%
  \BibitemOpen
  \bibfield  {author} {\bibinfo {author} {\bibfnamefont {N.~D.}\ \bibnamefont
  {Birrell}}\ and\ \bibinfo {author} {\bibfnamefont {P.~C.~W.}\ \bibnamefont
  {Davies}},\ }\href {https://doi.org/10.1017/CBO9780511622632} {\emph
  {\bibinfo {title} {Quantum Fields in Curved Space}}},\ Cambridge Monographs
  on Mathematical Physics\ (\bibinfo  {publisher} {Cambridge University
  Press},\ \bibinfo {year} {1982})\BibitemShut {NoStop}%
\bibitem [{\citenamefont {Fedotov}\ \emph {et~al.}(2011)\citenamefont
  {Fedotov}, \citenamefont {Gelfer}, \citenamefont {Korolev},\ and\
  \citenamefont {Smolyansky}}]{Fedotov_2011}%
  \BibitemOpen
  \bibfield  {author} {\bibinfo {author} {\bibfnamefont {A.~M.}\ \bibnamefont
  {Fedotov}}, \bibinfo {author} {\bibfnamefont {E.~G.}\ \bibnamefont {Gelfer}},
  \bibinfo {author} {\bibfnamefont {K.~Y.}\ \bibnamefont {Korolev}},\ and\
  \bibinfo {author} {\bibfnamefont {S.~A.}\ \bibnamefont {Smolyansky}},\
  }\bibfield  {title} {\bibinfo {title} {{Kinetic equation approach to pair
  production by a time-dependent electric field}},\ }\href@noop {} {\bibfield
  {journal} {\bibinfo  {journal} {Phys. Rev. D}\ }\textbf {\bibinfo {volume}
  {83}},\ \bibinfo {pages} {025011} (\bibinfo {year} {2011})}\BibitemShut
  {NoStop}%
\bibitem [{\citenamefont {Parker}\ and\ \citenamefont
  {Toms}(2009)}]{parker_toms_2009}%
  \BibitemOpen
  \bibfield  {author} {\bibinfo {author} {\bibfnamefont {L.}~\bibnamefont
  {Parker}}\ and\ \bibinfo {author} {\bibfnamefont {D.}~\bibnamefont {Toms}},\
  }\href {https://doi.org/10.1017/CBO9780511813924} {\emph {\bibinfo {title}
  {Quantum Field Theory in Curved Spacetime: Quantized Fields and Gravity}}},\
  Cambridge Monographs on Mathematical Physics\ (\bibinfo  {publisher}
  {Cambridge University Press},\ \bibinfo {year} {2009})\BibitemShut {NoStop}%
\bibitem [{\citenamefont {Taya}\ \emph {et~al.}(2021)\citenamefont {Taya},
  \citenamefont {Fujimori}, \citenamefont {Misumi}, \citenamefont {Nitta},\
  and\ \citenamefont {Sakai}}]{exactWKB}%
  \BibitemOpen
  \bibfield  {author} {\bibinfo {author} {\bibfnamefont {H.}~\bibnamefont
  {Taya}}, \bibinfo {author} {\bibfnamefont {T.}~\bibnamefont {Fujimori}},
  \bibinfo {author} {\bibfnamefont {T.}~\bibnamefont {Misumi}}, \bibinfo
  {author} {\bibfnamefont {M.}~\bibnamefont {Nitta}},\ and\ \bibinfo {author}
  {\bibfnamefont {N.}~\bibnamefont {Sakai}},\ }\bibfield  {title} {\bibinfo
  {title} {{Exact WKB analysis of the vacuum pair production by time-dependent
  electric fields}},\ }\href {https://doi.org/10.1007/jhep03(2021)082}
  {\bibfield  {journal} {\bibinfo  {journal} {JHEP}\ }\textbf {\bibinfo
  {volume} {2021}},\ \bibinfo {pages} {82 (2021)}}\BibitemShut {NoStop}%
\bibitem [{\citenamefont {Beltr^^c3^^a1n-Palau}\ \emph
  {et~al.}(2019)\citenamefont {Beltr^^c3^^a1n-Palau}, \citenamefont {Ferreiro},
  \citenamefont {Navarro-Salas},\ and\ \citenamefont
  {Pla}}]{BreakingAdiabaticNavarro}%
  \BibitemOpen
  \bibfield  {author} {\bibinfo {author} {\bibfnamefont {P.}~\bibnamefont
  {Beltr^^c3^^a1n-Palau}}, \bibinfo {author} {\bibfnamefont {A.}~\bibnamefont
  {Ferreiro}}, \bibinfo {author} {\bibfnamefont {J.}~\bibnamefont
  {Navarro-Salas}},\ and\ \bibinfo {author} {\bibfnamefont {S.}~\bibnamefont
  {Pla}},\ }\bibfield  {title} {\bibinfo {title} {Breaking of adiabatic
  invariance in the creation of particles by electromagnetic backgrounds},\
  }\href {https://doi.org/10.1103/physrevd.100.085014} {\bibfield  {journal}
  {\bibinfo  {journal} {Phys. Rev. D}\ }\textbf {\bibinfo {volume} {100}},\
  \bibinfo {pages} {085014} (\bibinfo {year} {2019})}\BibitemShut {NoStop}%
\bibitem [{\citenamefont {Schmidt}\ \emph {et~al.}(1999)\citenamefont
  {Schmidt}, \citenamefont {Blaschke}, \citenamefont {R^^c3^^b6pke},
  \citenamefont {Prozorkevich}, \citenamefont {Smolyansky},\ and\ \citenamefont
  {Toneev}}]{SchmidtNonMarkovian}%
  \BibitemOpen
  \bibfield  {author} {\bibinfo {author} {\bibfnamefont {S.}~\bibnamefont
  {Schmidt}}, \bibinfo {author} {\bibfnamefont {D.}~\bibnamefont {Blaschke}},
  \bibinfo {author} {\bibfnamefont {G.}~\bibnamefont {R^^c3^^b6pke}}, \bibinfo
  {author} {\bibfnamefont {A.}~\bibnamefont {Prozorkevich}}, \bibinfo {author}
  {\bibfnamefont {S.}~\bibnamefont {Smolyansky}},\ and\ \bibinfo {author}
  {\bibfnamefont {V.}~\bibnamefont {Toneev}},\ }\bibfield  {title} {\bibinfo
  {title} {{Non-Markovian effects in strong-field pair creation}},\ }\href
  {https://doi.org/10.1103/physrevd.59.094005} {\bibfield  {journal} {\bibinfo
  {journal} {Phys. Rev. D}\ }\textbf {\bibinfo {volume} {59}},\ \bibinfo
  {pages} {094005} (\bibinfo {year} {1999})}\BibitemShut {NoStop}%
\bibitem [{\citenamefont {Bloch}(1999)}]{Bloch1999}%
  \BibitemOpen
  \bibfield  {author} {\bibinfo {author} {\bibfnamefont {J.~C. R. e.~a.}\
  \bibnamefont {Bloch}},\ }\bibfield  {title} {\bibinfo {title} {{Pair
  creation: Back reactions and damping}},\ }\bibfield  {journal} {\bibinfo
  {journal} {Physical Review D}\ }\textbf {\bibinfo {volume} {60}},\ \href
  {https://doi.org/10.1103/physrevd.60.116011} {10.1103/physrevd.60.116011}
  (\bibinfo {year} {1999})\BibitemShut {NoStop}%
\bibitem [{\citenamefont {Shale}(1962)}]{Shale:1962}%
  \BibitemOpen
  \bibfield  {author} {\bibinfo {author} {\bibfnamefont {D.}~\bibnamefont
  {Shale}},\ }\bibfield  {title} {\bibinfo {title} {{Linear symmetries of free
  boson fields}},\ }\href {https://doi.org/10.1090/S0002-9947-1962-0137504-6}
  {\bibfield  {journal} {\bibinfo  {journal} {Trans. Am. Math. Soc.}\ }\textbf
  {\bibinfo {volume} {103}},\ \bibinfo {pages} {149} (\bibinfo {year}
  {1962})}\BibitemShut {NoStop}%
\bibitem [{\citenamefont {Ruijsenaars}(1978)}]{RUIJSENAARS1978105}%
  \BibitemOpen
  \bibfield  {author} {\bibinfo {author} {\bibfnamefont {S.}~\bibnamefont
  {Ruijsenaars}},\ }\bibfield  {title} {\bibinfo {title} {{On Bogoliubov
  transformations. 2. The general case}},\ }\href
  {https://doi.org/https://doi.org/10.1016/0003-4916(78)90006-4} {\bibfield
  {journal} {\bibinfo  {journal} {Ann. Phys.}\ }\textbf {\bibinfo {volume}
  {116}},\ \bibinfo {pages} {105} (\bibinfo {year} {1978})}\BibitemShut
  {NoStop}%
\bibitem [{\citenamefont {Kim}\ and\ \citenamefont
  {Schubert}(2011)}]{Nonadiabatic_Kim}%
  \BibitemOpen
  \bibfield  {author} {\bibinfo {author} {\bibfnamefont {S.~P.}\ \bibnamefont
  {Kim}}\ and\ \bibinfo {author} {\bibfnamefont {C.}~\bibnamefont {Schubert}},\
  }\bibfield  {title} {\bibinfo {title} {{Nonadiabatic quantum Vlasov equation
  for Schwinger pair production}},\ }\href
  {https://doi.org/10.1103/physrevd.84.125028} {\bibfield  {journal} {\bibinfo
  {journal} {Phys. Rev. D}\ }\textbf {\bibinfo {volume} {84}},\ \bibinfo
  {pages} {125028} (\bibinfo {year} {2011})}\BibitemShut {NoStop}%
\bibitem [{\citenamefont {Huet}\ \emph {et~al.}(2014)\citenamefont {Huet},
  \citenamefont {Kim},\ and\ \citenamefont {Schubert}}]{Huet_2014}%
  \BibitemOpen
  \bibfield  {author} {\bibinfo {author} {\bibfnamefont {A.}~\bibnamefont
  {Huet}}, \bibinfo {author} {\bibfnamefont {S.~P.}\ \bibnamefont {Kim}},\ and\
  \bibinfo {author} {\bibfnamefont {C.}~\bibnamefont {Schubert}},\ }\bibfield
  {title} {\bibinfo {title} {{Vlasov equation for Schwinger pair production in
  a time-dependent electric field}},\ }\href
  {https://doi.org/10.1103/physrevd.90.125033} {\bibfield  {journal} {\bibinfo
  {journal} {Phys. Rev. D}\ }\textbf {\bibinfo {volume} {90}},\ \bibinfo
  {pages} {125033} (\bibinfo {year} {2014})}\BibitemShut {NoStop}%
\bibitem [{\citenamefont {Ruijsenaars}(1977)}]{Ruijsencharged}%
  \BibitemOpen
  \bibfield  {author} {\bibinfo {author} {\bibfnamefont {S.~N.~M.}\
  \bibnamefont {Ruijsenaars}},\ }\bibfield  {title} {\bibinfo {title} {{Charged
  particles in external fields. I. Classical theory}},\ }\href
  {https://doi.org/10.1063/1.523334} {\bibfield  {journal} {\bibinfo  {journal}
  {J. Math. Phys.}\ }\textbf {\bibinfo {volume} {18}},\ \bibinfo {pages} {720}
  (\bibinfo {year} {1977})}\BibitemShut {NoStop}%
\bibitem [{\citenamefont {Parker}\ and\ \citenamefont
  {Fulling}(1974)}]{PhysRevD.9.341}%
  \BibitemOpen
  \bibfield  {author} {\bibinfo {author} {\bibfnamefont {L.}~\bibnamefont
  {Parker}}\ and\ \bibinfo {author} {\bibfnamefont {S.~A.}\ \bibnamefont
  {Fulling}},\ }\bibfield  {title} {\bibinfo {title} {Adiabatic regularization
  of the energy-momentum tensor of a quantized field in homogeneous spaces},\
  }\href {https://doi.org/10.1103/PhysRevD.9.341} {\bibfield  {journal}
  {\bibinfo  {journal} {Phys. Rev. D}\ }\textbf {\bibinfo {volume} {9}},\
  \bibinfo {pages} {341} (\bibinfo {year} {1974})}\BibitemShut {NoStop}%
\end{thebibliography}%

\end{document}